\documentclass[showpacs,twocolumn,nofootinbib]{revtex4}
\usepackage{amsfonts}
\usepackage{makeidx}
\usepackage{amsmath}
\usepackage{graphicx}
\usepackage{amssymb}

\begin{document}

\title{The typical mass ratio and typical final spin in supermassive black
hole mergers}
\author{ L\'{a}szl\'{o} \'{A}. Gergely$^{1,2\dag }$, Peter L. Biermann$%
^{3,4,5,6,7\ddag }$}
\affiliation{$^{1}$ Department\ of Theoretical Physics, University\ of Szeged, Hungary\\
$^{2}$ Department\ of Experimental Physics, University\ of Szeged, Hungary\\
$^{3}~$Max-Planck-Institut f{\"{u}}r Radioastronomie, Auf dem H{\"{u}}gel
69, Bonn, Germany\\
$^{4}~$FZ Karlsruhe and Physics Department, University of Karlsruhe, Germany%
\\
$^{5}~$Department of Physics and Astronomy, University of Alabama,
Tuscaloosa, AL, USA \\
$^{6}~$Department of Physics, University of Alabama at Huntsville, AL, USA\\
$^{7}~$Department of Physics and Astronomy, University of Bonn, Germany\\
{\small $^{\dag }$ E-mail: gergely@physx.u-szeged.hu \qquad $^{\ddag }$
E-mail: plbiermann@mpifr-bonn.mpg.de}}

\begin{abstract}
We prove that merging supermassive black holes (SMBHs) typically have
neither equal masses, nor is their mass ratio too extreme. The majority of
such mergers fall into the mass ratio range of $1:30$ to $1:3$, implying a
spin flip during the inspiral. We also present a simple expression for the
final spin $\chi _{f}$ of the emerging SMBH, as function of the mass ratio,
initial spin magnitudes, and orientation of the spins with respect to the
orbital plane and each other. This formula approximates well more cumbersome
expressions obtained from the fit with numerical simulations. By integrating
over all equally likely orientations for precessing mergers we determine a
lower approximant to the final spin distribution as function of the mass
ratio alone. By folding this with the derived mass ratio dependent merger
rate we derive a \textit{lower} bound to the typical final spin value after
mergers. We repeat the procedure deriving an \textit{upper} bound for the
typical spin in the case when the spins are aligned to the orbital angular
momentum, such that there is no precession in the system. Both slopes of $%
\chi _{f}$ as function of the initial spins being smaller than one lead to
two attractors at $\chi _{f}^{prec}=0.2$ and $\chi _{f}^{align}=0.45$,
respectively. Real mergers, biased toward partial alignment by interactions
with the environment (accretion, host galaxy, etc.) would generate a typical
final spin lying between these two limiting values. These are the typical
values of the spin after the merger, starting from which the spin can built
up by further gaseous accretion.
\end{abstract}

\maketitle

\section{Introduction}

Einstein's general theory of relativity predicts that the coalescence of two
compact objects (neutron stars or black holes) is accompanied (and driven
by) intense gravitational radiation. Stellar mass (a few to a few ten solar
masses, M$_{\odot }$) black hole binaries emit gravitational waves with
frequency falling into the best sensitivity range of LIGO \cite{LIGO}, Virgo 
\cite{Virgo} and GEO600 \cite{GEO} Earth-based interferometric gravitational
wave detectors. Up to date there are very few observations \cite{IMBH}
indicating the existence of intermediate mass black holes. Binaries formed
by such black holes would emit gravitational waves falling into the
frequency range of third generation gravitational wave detectors, like the
Einstein Telescope \cite{ET}.

Supermassive black holes (SMBHs) with masses of $10^{6}\div 3\times 10^{9}$ M%
$_{\odot }$ (or perhaps even higher) on the other hand are quite frequent,
residing in the centre of each sufficiently massive galaxy. Their growth
occurs by accretion phases and by mergers, the estimated contribution of
each of these processes to the growth of mass being model-dependent. Recent
observations \cite{Kormendy} show that some galaxies never merge, and yet
may have central back holes; representing either the pure SMBH birth
population, or the birth population with some gaseous accretion.

The accretion process has been modeled with the inclusion of magnetic
fields, electromagnetic radiation of the disk and energetic jets
transporting angular momentum from the polar regions \cite{Bardeen}-\cite%
{accretion}. Beside increasing the mass, accretion will also spin up the
black holes. The spin limit reached due to accretion by canonical black
holes (the system of a black hole and electromagnetically radiating
accretion disk), as expressed in terms of the dimensionless spin is $\chi
_{can}=0.998$, close to the theoretically allowed maximum for a black hole,
the unity. Indeed, such a high spin powering the jets seems compulsory for\
understanding the low energy cutoff in the energetic electron spectra of
jets in radio galaxies \cite{cutoff}. Active Galactic Nuclei, in particular
the closest, Cen A are the most likely sources of for the Ultra High Energy
Cosmic Rays \cite{UHECR}.

When galaxies merge, eventually their central SMBHs will also do so.
Dynamical friction transfers some of the orbital angular momentum of the
binary black hole system to the stellar environment, being ejected at the
poles, a process which drives the system through the last parsec \cite{Zier}%
. Other mechanisms to overcome the last parsec are relaxation processes due
to cloud/star -- star interactions, repopulating the stellar orbits in the
center of the galaxy \cite{Alexander}, binary orbital decay by three-body
interactions in the gravitationally bound stellar cusps \cite{Sesana}, or
the interplay of three accretion disks: one around each black hole and the
third, circumbinary, removing orbital angular momentum from the binary \cite%
{Hayasaki}.

At about $0.005$ parsecs gravitational radiation takes over dynamical
friction as the leading dissipative effect \cite{SpinFlip1}. For many SMBH
binaries the gravitational waves emitted in the process of coalescence fall
into the frequency range the long-delayed space mission LISA \cite{LISA}.
Depending on how rich in gas the binary environment may be and whether there
any circumbinary disk has been formed, certain alignment between the proper
spins and the orbital angular momentum could occur due to the
Bardeen-Petterson effect (based in turn on the Lense-Thirring precession) 
\cite{BaPe}. The two situations which could occur are the mergers precessing
under random angle (also known as dry mergers) and non-precessing mergers,
implying complete alignment of the spins and orbital angular momentum (wet
mergers). The randomness in the orientation of precessing mergers typically
reduces the final spin \cite{HuBl}. For equal mass precessing mergers this
varies from $0.69$ for non-spinning black holes up to values of $0.73$ for
maximally spinning black holes \cite{BertiVolonteri}. For a mass ratio $1:10$
the range of final spins opens up to the interval between $0.2$ and $0.83$,
respectively, as function of the initial spins. For non-precessing mergers
all configurations would practically conserve a high initial spin during the
inspiral.

When the two black holes are at large distances, the orbital angular
momentum $L$ is always much larger then the individual spins $S_{i}$.
However at the characteristic radius $r^{\ast }\approx 0.005$ parsec%
\footnote{%
The distance $r^{\ast }$ depends weakly, as $m^{5/11}$ on the total mass and
negligibly, as $\eta ^{2/11}$ on the symmetric mass ratio $\eta =\mu
/m=\left( q^{1/2}+q^{-1/2}\right) ^{-2}$, where $\mu =m_{1}m_{2}/m$ is the
reduced mass.} (and the corresponding post-Newtonian (PN) parameter $%
\varepsilon ^{\ast }=Gm/c^{2}r^{\ast }\approx 10^{-3}$, defining the
beginning of the inspiral, see \cite{SpinFlip1}), where gravitational
radiation takes over dynamical friction as the leading order dissipative
effect, the ratios $S_{i}/L\approx \left( \varepsilon ^{\ast }\right)
^{1/2}q^{3-2i}\chi _{i}$ depend on the mass ratio $q\geq 1$. For a maximally
spinning larger black hole and separation $r_{\ast }$ the ratio is one at
about $q\approx 30$. For mass ratios larger than $30$ therefore the spin
dominates over the orbital angular momentum during the whole inspiral.

During the inspiral gravitational radiation further reduces the orbital
angular momentum, but not the spin magnitudes. The spins will only precess
driven by the leading order spin-orbit coupling and corrections due to
spin-spin and mass quadrupole - mass monopole coupling \cite{BOC}. In the
process the direction of the total angular momentum remains unchanged, in an
averaged sense over one radial orbit \cite{ACST}.

Gravitational radiation does not modify this conclusion on short
time-scales. Radiative evolutions with spin-orbit \cite{GPV3}, spin-spin 
\cite{spinspin1} and mass quadrupole - mass monopole \cite{quadrup}
couplings have been investigated, and their analysis in Ref. \cite{spinspin2}
lead to the important result that the instantaneous radiative changes of the
spins average out during a radial period. On this timescale therefore there
is no secular radiative change of the spin vectors at all:%
\begin{equation}
\left\langle \frac{d\mathbf{S_{i}}}{dt}\right\rangle =0\ .  \label{nospinave}
\end{equation}%
This result confirms that the spin dynamics can be regarded as a pure
precession, up to high PN orders including radiation reaction, on the
timescales comparable with a radial orbit. On much larger timescales however
the spin vector will undergo a reorientation (spin-flip), as explained in
detail in Ref. \cite{SpinFlip1}.

There are several scenarios possible, according to the actual mass ratio::

a) The masses are comparable $m_{2}\approx m_{1}$. In this case at the end
of the inspiral (when the PN approximation breaks down) the orbital angular
momentum still dominates over the spins \cite{SpinFlip1}, \cite{SpinFlip2}.
Radiating away this remnant orbital angular momentum during the merger
phase, while extrapolating the conservation of the direction of the total
angular momentum and of the individual spin magnitudes to this phase \cite%
{finalspin}, \cite{BR} could significantly reduce the final spin in all
cases when the individual spins were severely misaligned with each other and
with the orbital angular momentum. Such a misalignment would be typical in
the case of precessing mergers. Therefore for equal masses in a precessing
merger a not too high final spin can be considered typical.

b) The mass ratio is in the range $1:30$ to $1:3$. In this case the orbital
angular momentum dominates over the spin only at the beginning of the
inspiral, and as such is roughly aligned with the total angular momentum. At
the end of the inspiral however the orbital angular momentum becomes smaller
than the dominant spin, which has therefore to be reoriented towards the
invariant total angular momentum direction. For precessing mergers this
process causes a spin-flip during the inspiral, but does not reduce
significantly the magnitude of the dominant spin \cite{SpinFlip1}, \cite%
{SpinFlip2}. Non-precessing mergers on the other hand already imply an
alignment of the spins and orbital angular momentum, therefore neither the
spin magnitude, nor its direction will be changed by this process. In both
cases, whatever happens to the orbital angular momentum during the plunge,
its small value (compared to the dominant spin) at the end of the inspiral
will obstruct any serious further change in the final spin.

In this mass ratio range therefore the magnitude of the dominant spin will
not be much reduced by the merger, nevertheless a significant reorientation
of its direction during the inspiral will typically occur for precessing
mergers, which could be followed only by a minor further spin-flip during
the plunge.

c) The mass ratio is less than $1:30$. Then the orbital angular momentum is
too small from the beginning of the inspiral to modify the dominant spin.
Neither its magnitude, nor its direction are affected and we practically
face the inspiral of a test mass into the much larger black hole.

In this paper we revisit the merger process, based on the recent data of
Ref. \cite{CarameteBiermann}. We first derive the SMBH mass distribution. In
Section \ref{MassRatio} we fit a broken power law for the differential mass
function, then, based on this fit and a number of simple and reasonable
assumptions we derive the mass ratio distribution. We note that the results
of Ref. \cite{CarameteBiermann} are fully consistent with earlier results
based on much smaller statistics \cite{Greene}.

Next we derive in Section \ref{FinalSpin} a simple approximant for the final
spin of the emerging SMBH, as function of the mass ratio, initial spin
magnitudes, and orientation of the spins with respect to the orbital plane.
In the Appendix we compare the approximant with the more cumbersome
expressions existing in the literature, which were obtained by fit to
numerical simulations.

In Subsection \ref{Dry} we adopt the configuration of precessing mergers,
which allow for all relative spin and orbital angular momentum orientations
on equal footing, lowering the chances for a large final spin after the
merger. By integrating over all orientations in the precessing merger limit
(without allowing any preference for alignment), for any initial spin set we
determine a lower approximant to the final spin distribution as function of
the mass ratio alone. By folding this with the previously derived mass ratio
dependent merger rate, we obtain a \textit{lower bound to the typical final
spin} after SMBH mergers.

By contrast, in the non-precessing merger limit there is a perfect alignment
of the spins with the orbital angular momentum, hence the integration should
be carried on for this configuration alone, and only over the mass ratios,
folded with the mass ratio distribution. By this method, in Subsection \ref%
{Wet} we get an \textit{upper bound for the typical final spin}.

We discuss the implications of our results and present the concluding
remarks in Section \ref{Conclusions}.

\section{Mass ratios in SMBH mergers\label{MassRatio}}

In Ref. \cite{SpinFlip1} we gave a simple preliminary estimate of the
typical mass ratio of merging SMBHs. We revisit the problem more rigorously
here, both from a mathematical point of view and by employing new, more
precise data on SMBH masses, presented in Ref. \cite{CarameteBiermann}. We
do note that selection effects strongly influence some statistical
arguments, in the case, that selection is based on detectable activity at
the center of a galaxy for instance, on a far-infrared or ultra-violet
excess; in the first case this could be due to selecting for central
emission lines, in the second due to a central star-burst, and in the third
to a visible central accretion disk. Our approach, taken in this paper, does
suffer from the selection effect, that the work done by \cite%
{CarameteBiermann} used the colors of an old stellar population as the
starting point, and then cut the sample to include only early Hubble type
galaxies. However, allowing for a sample of late Hubble type galaxies would
not increase the merger rate very much, since such galaxies usually suffer
few if any mergers \cite{Kormendy}.

\subsection{The differential mass function}

\begin{figure}[tbph]
\begin{center}
\centering\includegraphics[width=8.5cm]{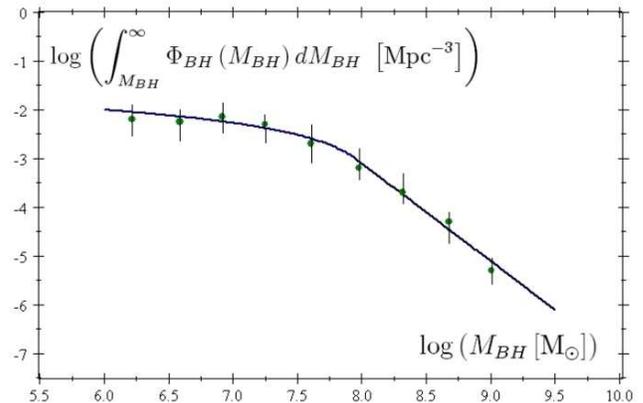}
\end{center}
\caption{(Color online) The integral mass function represented on
logarithmic scale shows a remarkably good fit to the data from Ref. 
\protect\cite{CarameteBiermann}. The differential mass function is taken as
a broken power law with powers $-1$ and $-3$, the breakpoint being at $%
8.9\times 10^{7}$M$_{\odot }$. }
\label{Fig_BH}
\end{figure}

The SMBH distribution $\Phi _{BH}\left( M_{BH}\right) \,$\ can be
interpreted as a power law with an exponential cutoff \cite{CarameteBiermann}%
. This can be well approximated by a broken power law \cite{PressSchechter}-%
\cite{Lauer}, also confirmed by the survey \cite{Ferrarese2}. The SMBH
integral mass function data represented on Fig 5. of Ref. \cite%
{CarameteBiermann}, after omitting the first two data points which do not
refer to black holes, but rather to nuclear star clusters \cite{Cote}, also
suggest the differential mass function $\Phi _{BH}(M_{BH})\propto M_{BH}^{-%
\tilde{\alpha}}$, with $\tilde{\alpha}=1$, starting from the lower mass
limit of $m_{a}\approx 10^{6}$ M$_{\odot }$ to the breakpoint, which is
approximately at $m_{\ast }\approx 10^{8}$ M$_{\odot }$; then $\Phi
_{BH}(M_{BH})\propto M_{BH}^{-\tilde{\beta}}$, with $\tilde{\beta}=3$,
starting from $m_{\ast }$ to the upper mass limit, taken here as $%
m_{b}\approx 3\times 10^{9}$ M$_{\odot }$. We prove this statement in the
remaining part of the subsection.\footnote{%
Note, that the lower mass data points were not important for the
considerations in Ref. \cite{CarameteBiermann}, concerned mainly with the
highest energy cosmic rays, such that a different fit of $\tilde{\alpha}%
_{CB}=2$ was advanced there. Nevertheless the supermassive black holes with
lower mass are important in the merger statistics, therefore in this paper
we chose $\tilde{\alpha}=1$ due to the tendency of the first data points to
be aligned horizontally (see Fig \ref{Fig_BH}). The limit $m_{a}$ is lowered
here as compared to the choice of Ref. \cite{SpinFlip1} such that the mass
of the SMBH in the centre of our Galaxy is not the lower mass limit any more.%
}

The SMBH data is represented on Fig \ref{Fig_BH}, which shows the integral
mass function $\int_{M_{BH}}^{\infty }\Phi \left( M_{BH}\right) dM_{BH}$ (in
Mpc$^{-3}$) as a function of the SMBH masses (in M$_{\odot }$), represented
on log-log scale. Due to the breakpoint (and by normalizing all masses to $%
m_{\star }$), the integral mass function for any $M_{BH}\leq m_{\ast }$ is%
\begin{eqnarray}
&&\int_{M_{BH}\leq m_{\ast }}^{\infty }\Phi \left( M_{BH}\right) dM_{BH} 
\notag \\
&=&k\int_{M_{BH}}^{m_{\ast }}\left( \frac{M_{BH}}{m_{\ast }}\right)
^{-1}dM_{BH}  \notag \\
&&+k\int_{m_{\ast }}^{\infty }\left( \frac{M_{BH}}{m_{\ast }}\right)
^{-3}dM_{BH}  \notag \\
&=&km_{\ast }\left. \ln \left( \frac{M_{BH}}{m_{\ast }}\right) \right\vert
_{M_{BH}}^{m_{\ast }}-\frac{1}{2}km_{\ast }\left. \left( \frac{M_{BH}}{%
m_{\ast }}\right) ^{-2}\right\vert _{m_{\ast }}^{\infty }  \notag \\
&=&km_{\ast }\left[ \frac{1}{2}-\ln \left( \frac{M_{BH}}{m_{\ast }}\right) %
\right] ~,~
\end{eqnarray}%
while for any $M_{BH}\geq m_{\ast }$ is, respectively 
\begin{eqnarray}
&&\int_{M_{BH}\geq m_{\ast }}^{\infty }\Phi \left( M_{BH}\right) dM_{BH} 
\notag \\
&=&k\int_{M_{BH}}^{\infty }\left( \frac{M_{BH}}{m_{\ast }}\right)
^{-3}dM_{BH}  \notag \\
&=&-\frac{1}{2}km_{\ast }\left. \left( \frac{M_{BH}}{m_{\ast }}\right)
^{-2}\right\vert _{M_{BH}}^{\infty }=\frac{1}{2}km_{\ast }\left( \frac{M_{BH}%
}{m_{\ast }}\right) ^{-2}~.~
\end{eqnarray}%
Here $k$ is a dimensional normalization constant. Both expressions reduce to 
$km_{\ast }/2$ at $M_{BH}=m_{\ast }$. Comparing with the data at $m_{\ast }$
allows to fix $\log \left( km_{\ast }/2\right) \approx -3$, thus $km_{\ast
}/2=10^{-3}$. Therefore%
\begin{eqnarray}
&&\log \int_{M_{BH}}^{\infty }\Phi \left( M_{BH}\right) dM_{BH}=-3  \notag \\
&&+\log \left( 
\begin{array}{cc}
1+4.6\left( \log m_{\ast }-x\right) & ,~\text{if }M_{BH}\leq m_{\ast } \\ 
10^{-2x}m_{\ast }^{2} & ,~\text{if }M_{BH}\geq m_{\ast }%
\end{array}%
\right) ~,
\end{eqnarray}%
where $x=\log M_{BH}$. Because $M_{BH}$ is given is solar masses, so is $%
m_{\ast }$. The broken power law with powers $-1$ and $-3$ gives the best
fit with the data by setting the breakpoint at $m_{\ast }=10^{7.95}$M$%
_{\odot }\approx \allowbreak 8.9\times 10^{7}$M$_{\odot }$, as seen on Fig %
\ref{Fig_BH}. (Note that the breakpoint turns out to be shifted as compared
with the number given in Ref. \cite{SpinFlip1}.) The fit is remarkable, the
sum of the squares of the deviances between the points and the function
values, divided by the square of the error bars is only $0.22$.

\subsection{SMBH mass ratio distribution}

\begin{figure}[tbph]
\begin{center}
\centering\includegraphics[width=7.5cm]{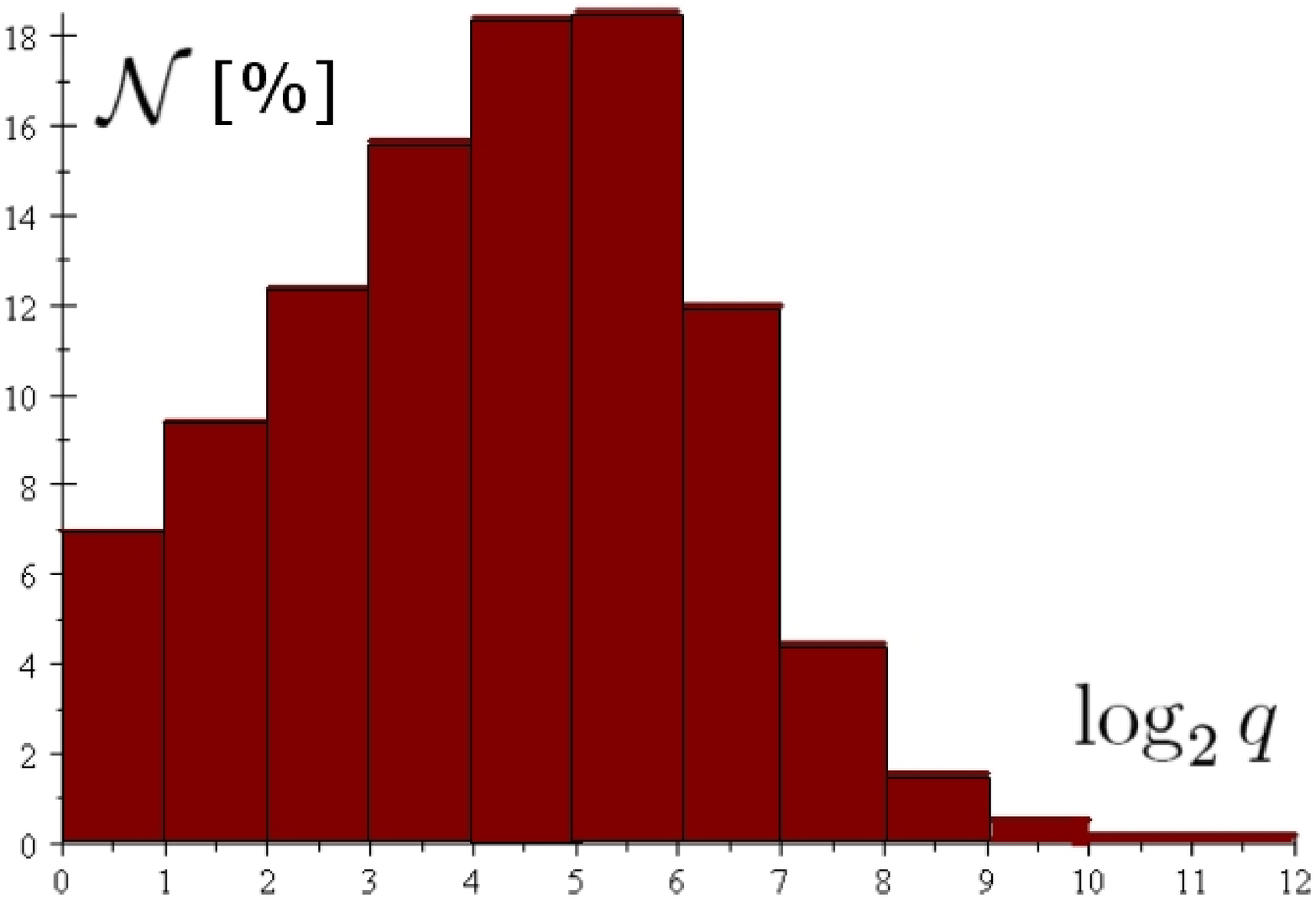}
\end{center}
\caption{(Color online) The number of SMBH encounters with mass ratios $q$
as function of $\log _{2}q$. }
\label{Fig_ratio}
\end{figure}

Based on the new SMBH mass function derived in the previous subsection here
we work out the estimates for the likelihood of the mass ratios, following
the logic of Ref. \cite{SpinFlip1}. However the changed mass values imply
more cases to be included in the analysis.

The number of encounters for a given mass ratio $q=m_{1}/m_{2}\geq 1$,
represented as $d\mathcal{N}/dq$ is proportional to the product of the
distribution functions for both black holes, folded with the merger
probability $F$ and integrated over the mass $m_{2}$ of the smaller black
hole:: 
\begin{equation}
\frac{d\mathcal{N}(q)}{dq}\propto \int_{m_{a}}^{m_{b}/q}\Phi
_{BH}(m_{2})\Phi _{BH}(qm_{2})F(q,m_{2})dm_{2}~.
\end{equation}%
The merger probability in turn is proportional to the cross section (we
neglect the weak dependence on the relative velocity of galaxies as these
are not too high, the Universe being not old enough for mass segregation).

In order to determine the cross section, we assume that each galaxy merger
is followed by the merger of their central SMBHs. Therefore we basically
evaluate the cross section of merging galaxies. Further, we note that the
masses of the galaxies and their central SMBHs correlate due to

\begin{itemize}
\item the correlation of the mass of the central SMBH with the mass of the
host galactic bulge \cite{Magorrian},

\item the proportionality of the mass of the central SMBH with both the
spheroidal galaxy mass component and the total mass (including dark matter)
of the galaxy \cite{Benson}.
\end{itemize}

It is likely that the more massive SMBH, thus the most massive galaxy
dominates the cross section, thus the merger rate $F$. As the cross section
is a function of the galaxy mass (thus SMBH mass), we take $F\sim \left(
qm_{2}\right) ^{\xi }$. We chose $\xi =1/2$ \ based on the following
observation:

\begin{itemize}
\item the comparison of our galaxy with dwarf spheroidals shows that an
increase by a factor of $10$ in radius (thus $10^{2}$ in cross section)\ is
accompanied by an increase by a factor of $10^{4}$ in mass \cite{Gilmore}-%
\cite{Klypin}.
\end{itemize}

The break point $m_{\star }$ splits the SMBH range into two intervals,
encompassing a mass range of about a factor of $q_{1}=89$ and $%
q_{2}=\allowbreak 36$. Thus (by normalizing all masses to $m_{\star }$) we
estimate for any $q\in \lbrack 1,36]$ the number of encounters as 
\begin{eqnarray}
&&\left. \frac{d\mathcal{N}(q)}{dq}\right. _{q\in \left[ 1,36\right] }\propto
\notag \\
&&\int_{m_{a}}^{m_{\star }/q}\left( \frac{m_{2}}{m_{\star }}\right) ^{-%
\tilde{\alpha}}\left( \frac{m_{2}q}{m_{\star }}\right) ^{-\tilde{\alpha}%
}\left( \frac{m_{2}q}{m_{\star }}\right) ^{\xi }dm_{2}  \notag \\
&+&\int_{m_{\star }/q}^{m_{\star }}\left( \frac{m_{2}}{m_{\star }}\right) ^{-%
\tilde{\alpha}}\left( \frac{m_{2}q}{m_{\star }}\right) ^{-\tilde{\beta}%
}\left( \frac{m_{2}q}{m_{\star }}\right) ^{\xi }dm_{2}  \notag \\
&+&\int_{m_{\star }}^{m_{b}/q}\left( \frac{m_{2}}{m_{\star }}\right) ^{-%
\tilde{\beta}}\left( \frac{m_{2}q}{m_{\star }}\right) ^{-\tilde{\beta}%
}\left( \frac{m_{2}q}{m_{\star }}\right) ^{\xi }dm_{2}~.  \label{N1}
\end{eqnarray}%
The first, second and third lines of the right hand side of Eq. (\ref{N1})
contain, respectively, the mergers of: two SMBHs from the lower mass
interval; one SMBH from the lower, the other from the higher mass interval;
and both SMBHs from the upper mass interval. The condition $q\leq q_{2}$
assures that the upper limit of the integrals is larger than the lower limit.

For $q\in \lbrack 36,89]$ there are two contributions, arising from the
combination of either two light or a light and a heavy SMBHs: 
\begin{eqnarray}
&&\left. \frac{d\mathcal{N}(q)}{dq}\right. _{q\in \left[ 36,89\right] }
\propto  \notag \\
&&\int_{m_{a}}^{m_{\star }/q}\left( \frac{m_{2}}{m_{\star }}\right) ^{-%
\tilde{\alpha}}\left( \frac{m_{2}q}{m_{\star }}\right) ^{-\tilde{\alpha}%
}\left( \frac{m_{2}q}{m_{\star }}\right) ^{\xi }dm_{2}  \notag \\
&&+\int_{m_{\star }/q}^{m_{b}/q}\left( \frac{m_{2}}{m_{\star }}\right) ^{-%
\tilde{\alpha}}\left( \frac{m_{2}q}{m_{\star }}\right) ^{-\tilde{\beta}%
}\left( \frac{m_{2}q}{m_{\star }}\right) ^{\xi }dm_{2}~.  \label{N2}
\end{eqnarray}%
The condition $q\leq q_{1}$ again assures that the upper limit of the
integrals is larger than the lower limit.

Finally for $q\in \lbrack 89,3000]$ there is one single contribution%
\begin{eqnarray}
&&\left. \frac{d\mathcal{N}(q)}{dq}\right. _{q\in \left[ 89,3000\right] }
\propto  \notag \\
&&\int_{m_{a}}^{m_{b}/q}\left( \frac{m_{2}}{m_{\star }}\right) ^{-\tilde{%
\alpha}}\left( \frac{m_{2}q}{m_{\star }}\right) ^{-\tilde{\beta}}\left( 
\frac{m_{2}q}{m_{\star }}\right) ^{\xi }dm_{2}~.  \label{N3}
\end{eqnarray}%
Eq. (\ref{N2}) expresses the encounters of a light SMBH from the lower
interval with a heavy SMBH from the upper interval.

Integration over $m_{2}$ gives 
\begin{gather}
\left. \frac{d\mathcal{N}(q)}{dq}\right. _{q\in \left[ 1,36\right] }\propto 
\frac{q^{-1+\tilde{\alpha}}-q_{1}^{-1-\xi +2\tilde{\alpha}}q^{\xi -\tilde{%
\alpha}}}{1+\xi -2\tilde{\alpha}}  \notag \\
+\frac{q^{\xi -\tilde{\beta}}-q^{-1+\tilde{\alpha}}}{1+\xi -\tilde{\alpha}-%
\tilde{\beta}}+\frac{q_{2}^{1+\xi -2\tilde{\beta}}q^{-1+\tilde{\beta}%
}-q^{\xi -\tilde{\beta}}}{1+\xi -2\tilde{\beta}}~,
\end{gather}%
\begin{gather}
\left. \frac{d\mathcal{N}(q)}{dq}\right. _{q\in \left[ 36,89\right] }\propto 
\frac{q^{-1+\tilde{\alpha}}-q_{1}^{-1-\xi +2\tilde{\alpha}}q^{\xi -\tilde{%
\alpha}}}{1+\xi -2\tilde{\alpha}}  \notag \\
+\frac{\left( q_{2}^{1+\xi -\tilde{\alpha}-\tilde{\beta}}-1\right) q^{-1+%
\tilde{\alpha}}}{1+\xi -\tilde{\alpha}-\tilde{\beta}}~.
\end{gather}%
and%
\begin{equation}
\left. \frac{d\mathcal{N}(q)}{dq}\right. _{q\in \left[ 89,3000\right]
}\propto \frac{q_{2}^{1+\xi -\tilde{\alpha}-\tilde{\beta}}q^{-1+\tilde{\alpha%
}}-q_{1}{}^{-1-\xi +\tilde{\alpha}+\tilde{\beta}}q^{\xi -\tilde{\beta}}}{%
1+\xi -\tilde{\alpha}-\tilde{\beta}}~.
\end{equation}%
(We have employed $m_{b}/m_{\star }=q_{2}$ and $m_{\star }/m_{a}=q_{1}$.)
With the preferred parameter values $\tilde{\alpha}=1,~\tilde{\beta}=3$ and $%
\xi =1/2,~q_{1}=89,~q_{2}=36$ the number of encounters for the three ranges
simplifies to%
\begin{gather}
\left. \frac{d\mathcal{N}(q)}{dq}\right. _{q\in \left[ 1,36\right] }=\frac{%
9.\,\allowbreak 396\,4\times 10^{-2}}{q^{0.5}}-\frac{8.\,\allowbreak
853\,6\times 10^{-4}}{q^{2.\,\allowbreak 5}}  \notag \\
-\allowbreak 1.\,\allowbreak 098\,2\times 10^{-10}q^{2}-7.\,\allowbreak
968\,1\times 10^{-3}~,  \notag \\
\left. \frac{d\mathcal{N}(q)}{dq}\right. _{q\in \left[ 36,89\right]
}=\allowbreak \frac{9.\,\allowbreak 396\,4\times 10^{-2}}{q^{0.5}}%
-7.\,\allowbreak 968\,6\times 10^{-3}  \notag \\
\left. \frac{d\mathcal{N}(q)}{dq}\right. _{q\in \left[ 89,3000\right] }=%
\frac{148.\,\allowbreak 86}{q^{2.\,\allowbreak 5}}-2.\,\allowbreak
561\,8\times 10^{-7}  \label{N}
\end{gather}%
Here we have normalized such that%
\begin{gather}
\int_{1}^{36}\left. \frac{d\mathcal{N}(q)}{dq}\right. _{q\in \left[ 1,36%
\right] }dq+\int_{36}^{89}\left. \frac{d\mathcal{N}(q)}{dq}\right. _{q\in %
\left[ 36,89\right] }dq  \notag \\
+\int_{89}^{3000}\left. \frac{d\mathcal{N}(q)}{dq}\right. _{q\in \left[
89,3000\right] }dq=1
\end{gather}%
holds.

Defining the number of encounters in a mass ratio interval $\left[
q_{1},q_{2}\right] $ as $\mathcal{N}_{q_{1}\div q_{2}}=\int_{q_{1}}^{q_{2}}%
\frac{d\mathcal{N}(q)}{dq}dq$ we obtain the percentages of the mergers with
the mass ratio ranges falling between $\left[ 1,3\right] $, $\left[ 3,30%
\right] $, $\left[ 30,100\right] $ and $\left[ 100,3000\right] $,
respectively as%
\begin{eqnarray}
\mathcal{N}_{1\div 3} &=&12.1~\%~,\quad \mathcal{N}_{3\div 30}=48.9~\%~, 
\notag \\
\mathcal{N}_{30\div 100} &=&29.2~\%,\quad \mathcal{N}_{100\div 3000}=9.8~\%~.
\end{eqnarray}%
$\allowbreak $The distribution of the mass ratios is shown in more detail on
the histogram of Fig \ref{Fig_ratio}.

The most likely mass ratio range, occurring in approximately half of the
mergers, turns out to be $q\in \left( 3\,,30\right) $, in agreement with the
rough estimate of Ref. \cite{SpinFlip1}. This is the mass ratio, where a
spin flip occurs during the inspiral \cite{SpinFlip1}. The second most
numerous mass ratio range, approximately in $30\%$ of the cases is for $q\in
\left( 30,100\right) $. Both the comparable mass case with $q\in \left(
1,3\right) $ and the extremal mass ratio case, defined here as $q\in \left(
100,3000\right) $ represent just about $10\%$ each of the SMBH mergers. This
important result makes compulsory to model SMBH mergers for non-equal masses.

\section{An approximate final spin formula in SMBH\ mergers\label{FinalSpin}}

In this section we propose a formula for the final spin, which on the one
hand approximates reasonably well more cumbersome expressions derived from
fits with numerical runs, on the other hand is simple enough to facilitate
the numerical integrations we will carry on in the remaining part of the
paper. In this section we use $\nu =q^{-1}$.

In the system with the Newtonian orbital angular momentum on the $z$-axis
and the periastron on the $x$-axis the spins are 
\begin{equation*}
\mathbf{S}_{\mathbf{i}}=\frac{G}{c}m\mu \nu ^{2i-3}\chi _{i}\left( \sin
\kappa _{i}\cos \zeta _{i},\sin \kappa _{i}\sin \zeta _{i},\cos \kappa
_{i}\right) ~.
\end{equation*}%
The magnitude of the total spin $\mathbf{S}=\mathbf{S}_{\mathbf{1}}+\mathbf{S%
}_{\mathbf{2}}$ is found from%
\begin{eqnarray*}
\mathbf{S}^{2} &=&\left[ \mathbf{S}_{\mathbf{1}}^{2}+2\mathbf{S}_{\mathbf{1}%
}\cdot \mathbf{S}_{\mathbf{2}}+\mathbf{S}_{\mathbf{2}}^{2}\right] ^{1/2} \\
&=&\frac{G}{c}m\mu \left[ \sum_{i=1,2}\left( \nu ^{2i-3}\chi _{i}\right)
^{2}+2\chi _{1}\chi _{2}\cos \gamma \right] ^{1/2}
\end{eqnarray*}%
with%
\begin{equation}
\cos \gamma =\cos \kappa _{1}\cos \kappa _{2}+\sin \kappa _{1}\sin \kappa
_{2}\cos \left( \zeta _{2}-\zeta _{1}\right)  \label{cosga}
\end{equation}%
while the orbital angular momentum (assuming circular orbits, thus $%
L_{N}=\mu rv$), to leading order can be written as 
\begin{equation*}
\mathbf{L}_{\mathbf{N}}=\frac{G}{c}m\mu \varepsilon ^{-1/2}\left(
0,0,1\right) ~,
\end{equation*}%
where $\varepsilon =Gm/c^{2}r=v^{2}/c^{2}$ is the post-Newtonian parameter.
(This parameter increases as the black holes approach each other.)

The dimensionless version $\mathfrak{J}=cJ/Gm\mu $ of the magnitude of the
total angular momentum reads%
\begin{eqnarray*}
\mathfrak{J} &=&\frac{c}{Gm\mu }\left[ \left( \mathbf{L}_{\mathbf{N}}+%
\mathbf{S}\right) \cdot \left( \mathbf{L}_{\mathbf{N}}+\mathbf{S}\right) %
\right] ^{1/2} \\
&=&\left[ \mathbf{L}_{\mathbf{N}}^{2}\mathbf{+2L}_{\mathbf{N}}\cdot \left( 
\mathbf{S}_{\mathbf{1}}\mathbf{+S}_{\mathbf{2}}\right) +\mathbf{S}^{2}\right]
^{1/2} \\
&=&\bigl[\varepsilon ^{-1}+2\varepsilon ^{-1/2}\sum_{i=1,2}\nu ^{2i-3}\chi
_{i}\cos \kappa _{i} \\
&&+\sum_{i=1,2}\left( \nu ^{2i-3}\chi _{i}\right) ^{2}+2\chi _{1}\chi
_{2}\cos \gamma \bigr]^{1/2}~.
\end{eqnarray*}%
The final spin magnitude is denoted 
\begin{equation*}
S_{f}=\frac{G}{c}m_{f}^{2}\chi _{f}~.
\end{equation*}

We identify as an upper limit for the final spin the magnitude of the total
angular momentum at the end of the inspiral, obtaining $\chi _{f}=\eta
\left( m/m_{f}\right) ^{2}\mathfrak{J}_{f}$. Here $\mathfrak{J}_{f}=%
\mathfrak{J}\left( \varepsilon =\varepsilon _{f}\right) $ and $\eta =\mu
/m=\nu \left( 1+\nu \right) ^{-2}$. By introducing the efficiency of mass
conversion into gravitational radiation as $\epsilon _{GW}=1-m_{f}/m$, we
can express $m/m_{f}=\left( 1-\epsilon _{GW}\right) ^{-1}$. Hence%
\begin{eqnarray}
\chi _{f} &=&\frac{\eta }{\left( 1-\epsilon _{GW}\right) ^{2}}\bigl[%
\varepsilon _{f}^{-1}+2\varepsilon _{f}^{-1/2}\sum_{i=1,2}\nu ^{2i-3}\chi
_{i}\cos \kappa _{i}  \notag \\
&&+\sum_{i=1,2}\left( \nu ^{2i-3}\chi _{i}\right) ^{2}+2\chi _{1}\chi
_{2}\cos \gamma \bigr]^{1/2}  \label{spinfin}
\end{eqnarray}%
We are interested in establishing a lower boundary for the value of the
final spin, thus we set the efficiency to zero. The maximal value of the
bracket for any given $\chi _{i}$ arises when the spins are aligned with the
orbital angular momentum:%
\begin{equation}
\chi _{f}^{\max }=\eta \mathfrak{J}_{f}^{\max }=\eta \left( \varepsilon
_{f}^{-1/2}+\nu ^{-1}\chi _{1}+\nu \chi _{2}\right)
\end{equation}%
With $\varepsilon _{f}^{-1/2}=2$ (at two Schwarzschild radii, this is the
radius of the innermost bound circular orbit in the Schwarzschild geometry)
and for maximal spins this gives $\chi _{f}^{\max }=1$, irrespective of the
actual value of $\nu $. Therefore we normalize $\chi _{f}$ by setting $%
\varepsilon _{f}^{-1/2}=2$ in Eq. (\ref{spinfin}), and obtain a very simple
expression for the final spin:%
\begin{eqnarray}
\chi _{f} &=&\frac{\nu }{\left( 1+\nu \right) ^{2}}\bigl[4+4\sum_{i=1,2}\nu
^{2i-3}\chi _{i}\cos \kappa _{i}  \notag \\
&&+\sum_{i=1,2}\left( \nu ^{2i-3}\chi _{i}\right) ^{2}+2\chi _{1}\chi
_{2}\cos \gamma \bigr]^{1/2}~.  \label{spinfinlow}
\end{eqnarray}%
When the mass ratio is extreme ($\nu \rightarrow 0$), Eq. (\ref{spinfinlow})
correctly reproduces $\chi _{f}=\chi _{1}$, a result to be expected from the
test particle limit. This final spin function qualitatively reproduces well
the more cumbersome final spin expressions found in the literature from fits
with numerical runs. In the Appendix we compare in detail the expression (%
\ref{spinfinlow}) with the one presented in Ref. \cite{BR}, finding that for
the largest part of the parameter space it slightly underestimates the final
spin.

\section{The typical final spin\label{typical}}

\subsection{Precessing (randomly oriented) mergers\label{Dry}}

In this subsection we discuss the typical spin in the merger of two black
holes by assuming generic precessing mergers. As this implies complete
randomness in the relative angular momenta orientations, we integrate the
final spin formula (\ref{spinfinlow}) over all possible orientations. Then
we weight this orientation independent, but still mass ratio dependent final
spin with the probabilities for a given mass ratio (\ref{N}) derived earlier
in Section \ref{MassRatio} and integrate over the mass ratios, obtaining a
typical final spin as function of initial spin magnitudes only. As the
integration over the mass ratio implies to integrate over $q$ (according to
the method of evaluating the merger rate), we will rewrite $\nu =q^{-1}$ in
all expressions.

\subsubsection{Mass ratio dependent typical final spin}

We first integrate the expression of the final spin (\ref{spinfinlow}) over
all spin directions. By adopting the precessing merger model, we allow for
random spin orientations. The assumption of randomness sets $\cos \kappa
_{i} $ and $\zeta _{i}$, the cosine of the spin polar angles and the spin
azimuthal angles as evenly distributed random variables. Instead of the
individual azimuthal angles, the combination $\gamma =\zeta _{2}-\zeta _{1}$
appearing in Eq. (\ref{spinfinlow}) and representing the relative spin
azimuthal angle will be randomized.

Integrating over all orientations (and properly normalizing by $8\pi $) we
find therefore a lower bound for the mass ratio dependent final spin as: 
\begin{equation}
\chi _{f}^{prec}\!\left( \chi _{i},q\right) \!=\!\frac{1}{8\pi }%
\!\!\int_{0}^{2\pi }\!\!\!\int_{-1}^{1}\!\!\int_{-1}^{1}\!\chi _{f}\left(
\chi _{i},q,\kappa _{i},\gamma \right) d\cos \!\kappa _{1}d\cos \!\kappa
_{2}d\gamma ~.  \label{spinfindry}
\end{equation}

\begin{widetext} 

\begin{figure}[tbph]
\includegraphics[width=8cm]{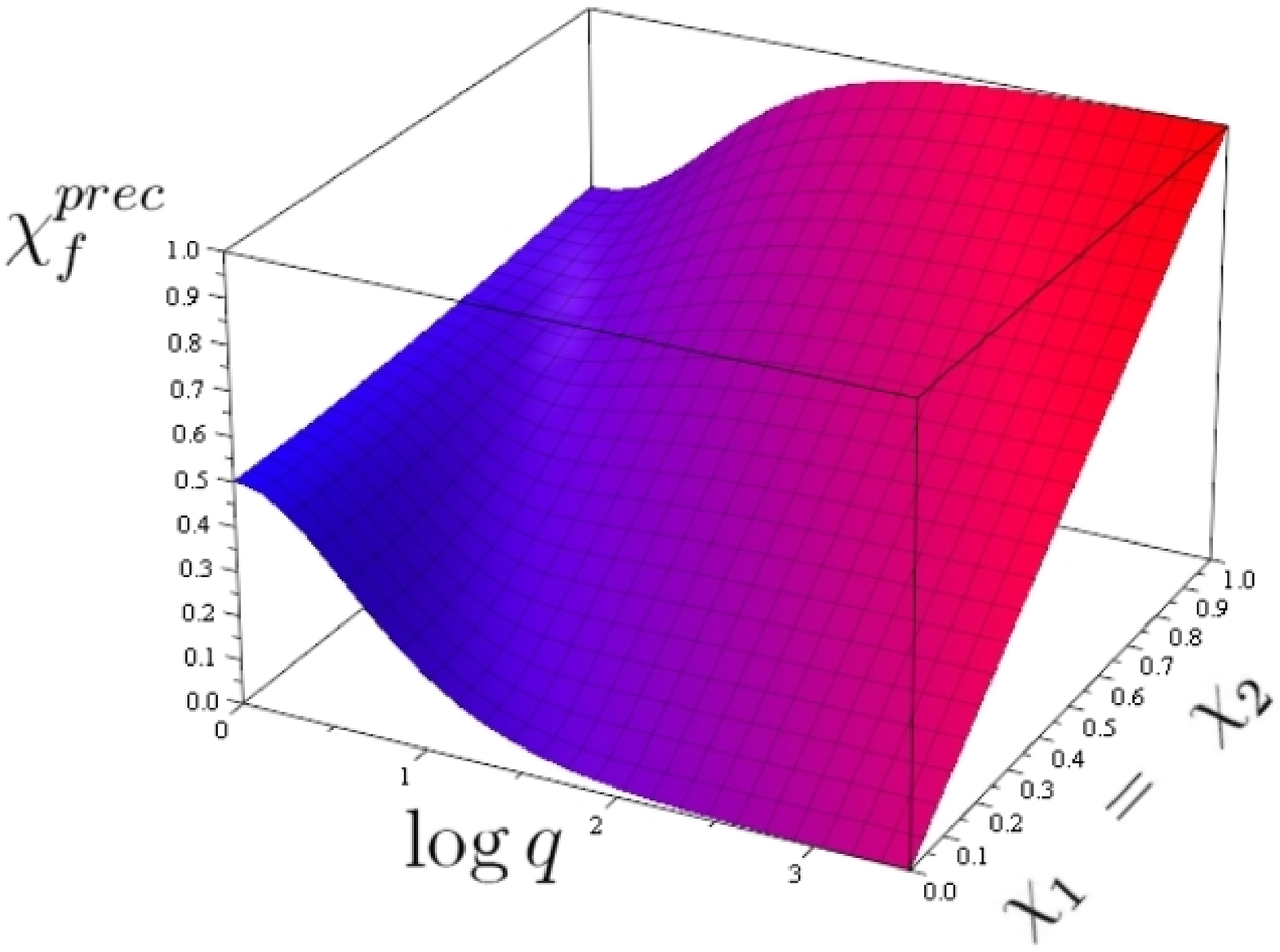}%
\includegraphics[width=8cm]{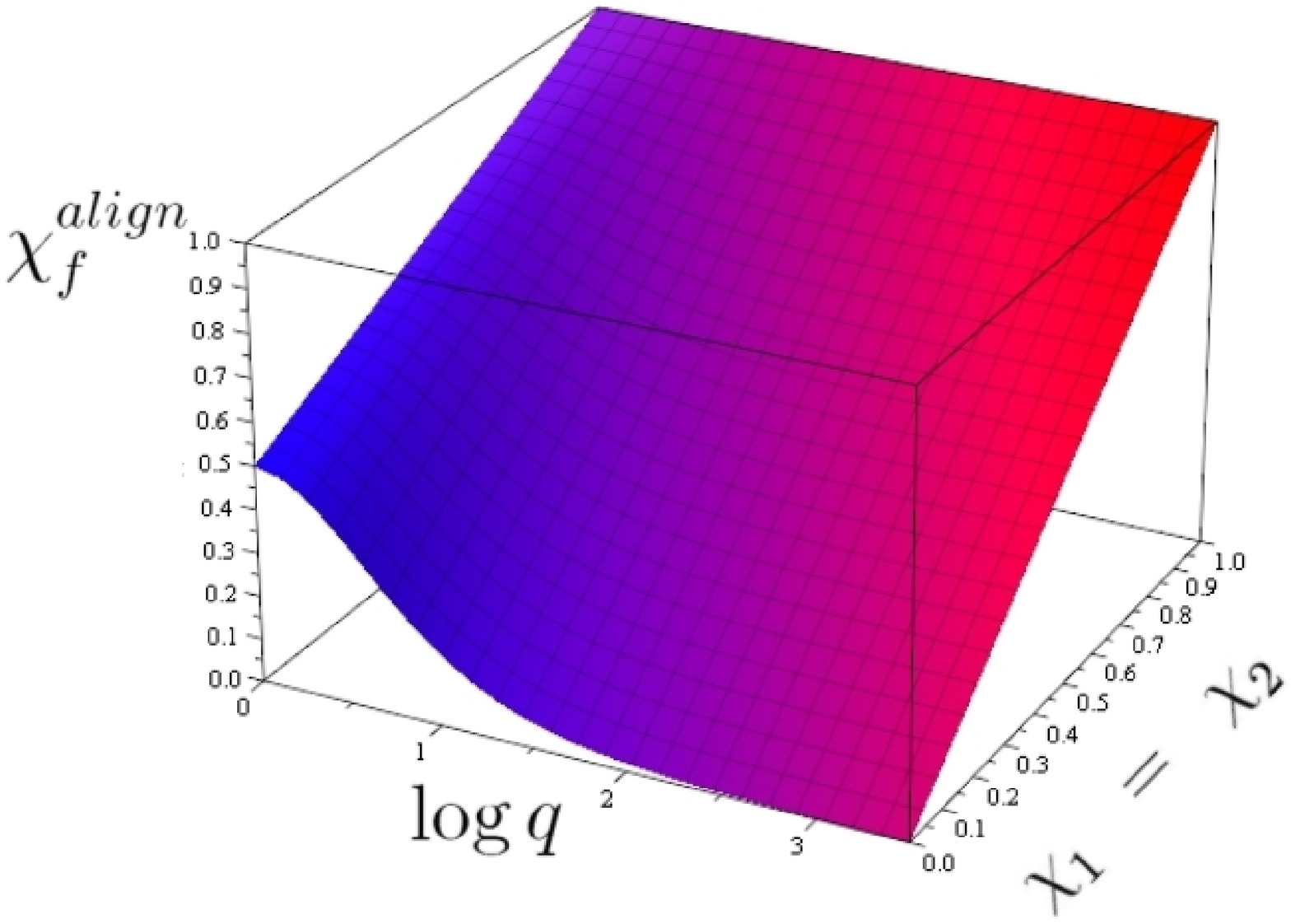}
\caption{(Color online) The typical final spin in supermassive black hole
mergers as function of $\protect\chi _{1}=\protect\chi _{2}$ and $\log q$,
represented for precessing mergers (averaged over random configurations) - left
panel; and mergers with the spins and orbital angular momentum fully
aligned - right panel.}
\label{Fig3}
\end{figure}

\end{widetext}

The final spin $\chi _{f}^{prec}$ for precessing mergers (random
configurations) as function of $\chi _{1}=\chi _{2}$ and $\log q$ is
represented in the left panel of Fig. \ref{Fig3}, as arising from a
numerical integration (with approximation by the midpoint method). For equal
masses the final spin ranges from $0.5\,\ $(for nonspinning black holes) to $%
0.6$ (for maximally spinning black holes). This is consistent with the
corresponding results of Ref. \cite{BertiVolonteri}. For $q\geq 100$ we have
the test particle limit: the final spin is accurately approximated by $\chi
_{1}$ (in other words the orbital angular momentum does not modify the spin
of the larger SMBH). In between there is the mass range with the most
frequent encounters. For $q\approx 10$\thinspace\ for example the lower
bound for the final spin ranges from $0.18$ (nonspinning mergers) to $0.85$
(maximally spinning mergers).

\subsubsection{Typical final spin in precessing mergers}

We establish an overall typical final spin for precessing mergers by
integrating Eq. (\ref{spinfindry}) over all possible mass ratios, weighted
with the mass ratio dependent probability of encounters given in Eqs. (\ref%
{N}):%
\begin{eqnarray}
\chi _{f}^{prec}\left( \chi _{i}\right) &=&\int_{1}^{36}\chi
_{f}^{prec}\left( \chi _{i},q\right) \left. \frac{d\mathcal{N}(q)}{dq}%
\right. _{q\in \left[ 1,36\right] }dq  \notag \\
&&+\int_{36}^{89}\chi _{f}^{prec}\left( \chi _{i},q\right) \left. \frac{d%
\mathcal{N}(q)}{dq}\right. _{q\in \left[ 36,89\right] }dq  \notag \\
&&+\int_{89}^{3000}\chi _{f}^{prec}\left( \chi _{i},q\right) \left. \frac{d%
\mathcal{N}(q)}{dq}\right. _{q\in \left[ 89,3000\right] }dq~.
\label{spinfintyp}
\end{eqnarray}%
The result of the numerical integration can be seen as the lower curve in
Fig \ref{Fig4}.

We note that for merging SMBHs in fast rotation $\chi _{1}=\chi _{2}\approx
0.998$ (the canonical spin limit, which occurs when both the accretion and
the radiation of the disk are taken into account \cite{PageThorne}) the
final spin becomes $\chi _{f}\approx 0.75$.

\subsection{Non-precessing (aligned) mergers\label{Wet}}

There is no precession in the perfectly aligned configurations, when the two
spins and the orbital angular momentum are parallel. Such configurations
could arise due to accretion or by other mechanisms. We do not model such
mechanisms here, just assume the alignment of the spins and orbital angular
momenta of the two-body system.

\subsubsection{The mass ratio dependent final spin}

The spins being aligned to the orbital angular momentum implies $\kappa
_{i}=0=\gamma $. Inserting these values in Eq. (\ref{spinfinlow}), the final
spin for non-precessing mergers as a function of the initial spin magnitudes
and mass ratio takes a remarkably simple form:%
\begin{equation}
\chi _{f}^{align}\left( \chi _{i},q\right) =\frac{q^{-1}}{\left(
1+q^{-1}\right) ^{2}}\left( 2+q\chi _{1}+q^{-1}\chi _{2}\right) ~.
\label{spinfinwet}
\end{equation}%
The final spin $\chi _{f}^{align}$ for non-precessing mergers is represented
in the right panel of Fig \ref{Fig3} as function of $\chi _{1}=\chi _{2}$
and $\log q$.

\subsubsection{Typical final spin in non-precessing mergers}

Next, we again integrate over the mass ratios, by properly weighting with
the mass ratio dependent probability of encounters, as given in Eqs. (\ref{N}%
):%
\begin{eqnarray}
\chi _{f}^{align}\left( \chi _{i}\right) &=&\int_{1}^{36}\chi
_{f}^{align}\left( \chi _{i},q\right) \left. \frac{d\mathcal{N}(q)}{dq}%
\right. _{q\in \left[ 1,36\right] }dq  \notag \\
&&+\int_{36}^{89}\chi _{f}^{align}\left( \chi _{i},q\right) \left. \frac{d%
\mathcal{N}(q)}{dq}\right. _{q\in \left[ 36,89\right] }dq  \notag \\
&&+\int_{89}^{3000}\chi _{f}^{align}\left( \chi _{i},q\right) \left. \frac{d%
\mathcal{N}(q)}{dq}\right. _{q\in \left[ 89,3000\right] }dq~.
\end{eqnarray}%
The result of the numerical integration can be seen as the upper curve in
Fig \ref{Fig4}.

For merging SMBHs in fast rotation $\chi _{1}=\chi _{2}\approx 0.998$ (the
canonical spin limit) the final spin is $\chi _{f}\approx 0.86$, much higher
than for precessing mergers, however still reduced essentially as compared
to the initial spin values.

\section{Concluding Remarks\label{Conclusions}}

In this paper we have studied the typical mass ratio and the typical final
spin in a two-body system composed of supermassive black holes (SMBH), thus
we did not consider the perturbations induced by either of the accretion
disks, nearby stellar population, magnetic fields or jets. SMBHs reside in
the center of each galaxy and following the frequent galaxy mergers they
also merge. Various dissipative processes, like dynamical friction,
accretion and emitted gravitational radiation are responsible to their
gradual approach and finally gravitational radiation is which drives them to
coalescence through a sequence of inspiral, merger and ringdown.

By starting from precise and new data on the SMBH mass distribution we
derived both a differential and an integral mass function, shown on Fig. \ref%
{Fig_BH}. The differential mass function is a broken power law, with
coefficients $\,-1$ and $-3$, with the breakpoint at $8.9\times 10^{7}$ M$%
_{\odot }$.

Then, exploiting a number of simple and reasonable assumptions we derived
the mass ratio dependent probability of encounters of two such SMBHs,
represented on Fig. \ref{Fig_ratio}. This confirms our expectation that the
most frequent (approximately half of the) encounters are for mass ratios $%
1:3 $ to $1:30$, the interesting mass ratio range where a spin-flip would
occur during the inspiral \cite{SpinFlip1}. 
\begin{figure}[tbph]
\begin{center}
\centering\includegraphics[width=8cm]{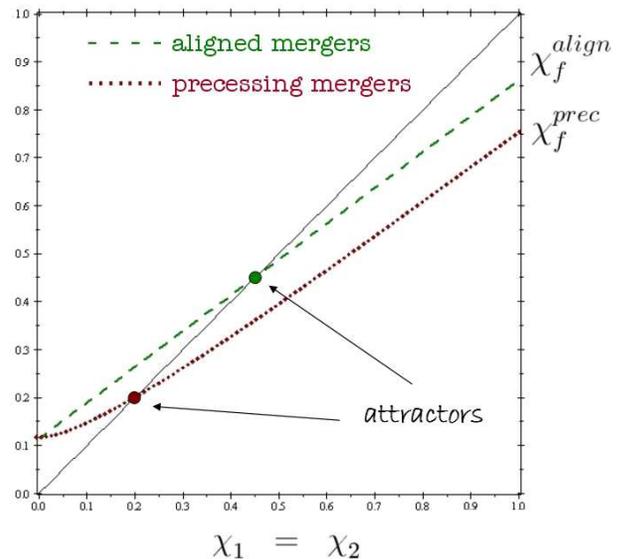}
\end{center}
\caption{(Color online) The typical final spin $\protect\chi _{f}$ as
function of $\protect\chi _{1}=\protect\chi _{2}$ only, in the randomly
precessing and the non-precessing merger limits (lower and upper curves,
respectively). The curves are obtained by integration over all mass ratios
of the expressions (\protect\ref{spinfindry}) $\protect\chi %
_{f}^{prec}\left( \protect\chi _{1}=\protect\chi _{2},q\right) $ and (%
\protect\ref{spinfinwet}) $\protect\chi _{f}^{align}\left( \protect\chi _{1}=%
\protect\chi _{2},q\right) $, respectively, weighted with the mass ratio
dependent probabilities of encounter (\protect\ref{N}). The line of equal
initial and final spins is also indicated. Where this line crosses the final
spin curves, there are two attractors (denoted by large dots), to where the
final spin would converge after a sequence of mergers in the two scenarios.}
\label{Fig4}
\end{figure}

Next, based on certain well-founded assumptions we derived a simple
analytical expression for the final spin of such a merger, depending on the
mass ratio, initial spin magnitudes, and orientation of the spins with
respect to the orbital plane and each other. This formula approximates well
more cumbersome expressions obtained from the fit with numerical
simulations, and it is much simpler, thus advantageous in order to carry on
the cumbersome numerical integrations which followed.

We proceeded to find the typical final spin in two limiting and highly
idealized scenarios. First we allowed for perfectly random orientations
(precessing case), over which we have integrated, obtaining a final spin
still depending on the initial spin magnitudes and mass ratio. Then we
folded with the derived mass ratio dependent merger rate, we integrated over
the mass ratio, deriving a lower bound to the typical final spin value after
mergers.

Secondly we considered the non-precessing configuration, with all spins and
the orbital angular momentum perfectly aligned. Folding the final spin for
this particular configuration again with the derived mass ratio dependent
merger rate and integrating over the mass ratio we obtained an upper bound
for the typical spin. These are represented as function of the initial spin
magnitudes (chosen to be equal\footnote{%
The second dimensionless spin parameter $\chi _{2}$ will anyhow have but a
small impact on the result, as the ratio of the \textit{total} spins $S_{i}$
scales with the mass ratio squared.}) on Fig. \ref{Fig4}. A third curve, the
line of equal initial and final spins is also indicated on the figure. The
fact that both slopes of $\chi _{f}$ as function of the initial spins are
smaller than one, leads to important consequences.

If we imagine a sequence of idealized (either randomly precessing or
non-precessing) mergers, what happens is that low spins tend to increase by
mergers while high spins decrease. There are in fact two attractors at $\chi
_{f}^{prec}=0.2$ and $\chi _{f}^{align}=0.45$, respectively, where the spins
converge after a reasonable number of the two types of mergers.

Real mergers, biased toward partial alignment by interactions with the
environment (accretion, host galaxy, etc.) would generate a typical final
spin lying between these two limiting values. Indeed, for example the galaxy
group distribution around NGC383 \footnote{This is also known as the 3C31 radio source, cf. the NASA Extragalactic Database.} looks like a spindle, with the spin of the
central black hole in the dominant galaxy, NGC383, aligned with the long
axis of the spindle. This shows a correlation between the dominant galaxy
black hole spin and the distribution of the other galaxies, all with central
black holes as well. It is to be expected that the distribution of galaxies
in the environment of a dominant galaxy is not random, but correlated, such
that in a merger of a galaxy with the dominant galaxy the final spin,
depending on the nature of the correlation, could fall anywhere between the
two curves shown of Fig. \ref{Fig4}.

After the merger episode gaseous accretion can start to increase the spin
again. If gaseous accretion were strong, then the spin could become quite
large in relatively short time. We propose to work out quantitatively such a
model in a forthcoming work.

\section{Acknowledgements}

This work has been realized in large part during two visits of L\'{A}G at
the Institute of Radioastronomy Bonn, supported through STSM grants of the
COST Action MP0905 "Black Holes in a Violent Universe".

\appendix

\section{Comparison of the final spin formula with related results}

Various papers have presented empirical formulae for the final spin, with
the functional form partially motivated by PN expressions and coefficients
fitted to the result of numerical runs. In one of the latest such works
Barausse and Rezzola \cite{BR} suggested a set of criteria I-V) such a
formula should obey. We compare this formula with our Eq. (\ref{spinfinlow}).

The condition I) of Ref. \cite{BR} implies no rest mass loss by
gravitational radiation; our lower spin limit estimate with $\epsilon
_{GW}=0 $ does the same. Condition III) assumes that the radiation loss in
the last stage of the merger is along the direction of the total angular
momentum, this property continuing to hold similarly as during the inspiral 
\cite{ACST}. By identifying $\chi _{f}$ with $\mathfrak{J}$ we also assume
that. Similarly, this identification assures the validity of condition IV),
which translates to no change in $\kappa _{i}$ and $\gamma $ during the
plunge. However we note that in the strict sense the spin-spin and
quadrupole-monopole couplings of the PN dynamics will obstruct this
assumption; therefore this assumption cannot be considered valid for any
distance, as assumed in Ref. \cite{BR}. Nevertheless these angular
evolutions are negligible on the orbital timescale, thus during the plunge
(over which we assume its validity and which lasts only from a fraction of
an orbit to a few orbits) the condition can be regarded as accurate.
Condition V. of Ref. \cite{BR} implies that the initial spins should drop
out completely from the final spin formula in the particular case of equal
masses and equal, but opposed spins. Our Eq. (\ref{spinfin}) can be
specified for this configuration by inserting $\nu =1,$ $\chi _{2}=\chi
_{1},~\cos \gamma =-1$ and $\cos \kappa _{2}=-\cos \kappa _{1}$ and it gives 
$\chi _{f}^{spec.config.}=\varepsilon _{f}^{-1/2}/4$, which is also
independent of the initial spins. Therefore condition V) also holds.

Finally, condition II) assumes that there are three vectors with conserved
length. These are the two spins (that we also assume), and the vector $%
\mathbf{L}_{\mathbf{N}}-\mathbf{J}+\mathbf{S}_{f}$ (what we do not). The
latter condition in our notation implies 
\begin{eqnarray}
\text{const.} &=&\left( \varepsilon ^{-1/2}\mathbf{\hat{L}}_{\mathbf{N}%
}-\left( \mathfrak{J}-\eta ^{-1}\chi _{f}\right) \mathbf{\hat{J}}\right) ^{2}
\notag \\
&=&\varepsilon ^{-1}+\left( \mathfrak{J}-\eta ^{-1}\chi _{f}\right) ^{2} 
\notag \\
&&-2\varepsilon ^{-1/2}\left( \mathfrak{J}-\eta ^{-1}\chi _{f}\right) 
\mathbf{\hat{L}}_{\mathbf{N}}\cdot \mathbf{\hat{J}~.}  \label{constr}
\end{eqnarray}%
We can set the constant to $\varepsilon _{f}^{-1}$ by evaluating the formula
at $\varepsilon _{f}$, where $\chi _{f}=\eta \mathfrak{J}_{f}$. Note that $%
\varepsilon ^{-1}\propto r$ thus it changes with $\dot{r}$, at Keplerian
order. For generic $r$ we have $\chi _{f}$=const. and $\mathfrak{J}$
changing significantly only on the radiation timescale (due to gravitational
radiation). Over the orbital timescale the change in $\mathfrak{J}$ by
gravitational radiation backreaction is at 2.5PN orders, while over the
precessional timescale is of 1PN order. The evolution of $\alpha =\cos
^{-1}\left( \mathbf{\hat{L}}_{\mathbf{N}}\cdot \mathbf{\hat{J}}\right) $
generates a 1PN change over the orbital timescale \cite{Inspiral}, therefore
the leading order changes over the orbital timescale of the second and third
terms in the second line on the right hand side of Eq. (\ref{constr}) are of
order $\varepsilon ^{5/2}\,$and $\varepsilon ^{1/2}$. Thus we conclude that
condition II) concerning the constancy of the length of the vector $\mathbf{L%
}_{\mathbf{N}}-\mathbf{J}+\mathbf{S}_{f}$ cannot be extended to arbitrary $r$%
, as in fact changes with $\varepsilon ^{1/2}$.

In the most generic case discussed in Ref. \cite{BR}, their Eqs. (6) and (8)
reproduce our Eq. (\ref{spinfin}), provided we replace their $\left\vert 
\mathfrak{l}\right\vert $, given by their Eq. (10) and rewritten in our
notations as%
\begin{eqnarray}
\left\vert \mathfrak{l}\right\vert  &=&2\sqrt{3}-3.5171\frac{\nu }{\left(
1+\nu \right) ^{2}}+2.5763\frac{\nu ^{2}}{\left( 1+\nu \right) ^{4}}  \notag
\\
&&+\frac{0.4537\frac{\nu }{\left( 1+\nu \right) ^{2}}-0.8904}{1+\nu ^{2}}%
\left( \chi _{1}\cos \kappa _{1}+\nu ^{2}\chi _{2}\cos \kappa _{2}\right)  
\notag \\
&&-\frac{0.1229}{\left( 1+\nu ^{2}\right) ^{2}}\left( \chi _{1}^{2}+\nu
^{4}\chi _{2}^{2}+2\nu ^{2}\chi _{1}\chi _{2}\cos \gamma \right) ~  \label{l}
\end{eqnarray}%
with $\varepsilon _{f}^{-1/2}$. In what follows, we compare the two values (%
\ref{spinfin}) for the final spin, once computed by replacing $\varepsilon
_{f}^{-1/2}$ with $\left\vert \mathfrak{l}\right\vert $ given by Eq. (\ref{l}%
), then with $\varepsilon _{f}^{-1/2}=2$. For example in the equal mass $\nu
=1$, equal spin $\chi _{1}=\chi _{2}$ case, when the spins are opposed to
each other, thus $\gamma =\pi $ and $\kappa _{2}=\pi -\kappa _{1}$, the
ratio $\chi _{f}^{BR}/\chi _{f}$ $\ $is identically $1.37$, regardless of
the values of $\chi _{1}=\chi _{2}$ and $\kappa _{1}$, therefore $\chi _{f}$
underestimates $\chi _{f}^{BR}$. Various other configurations, all for equal
dimensionless spins $\chi _{2}=\chi _{1}$, are represented on Figs. \ref%
{Fig1} and \ref{Fig2}.

\begin{widetext}

\begin{figure}[ht]
\includegraphics[width=7cm]{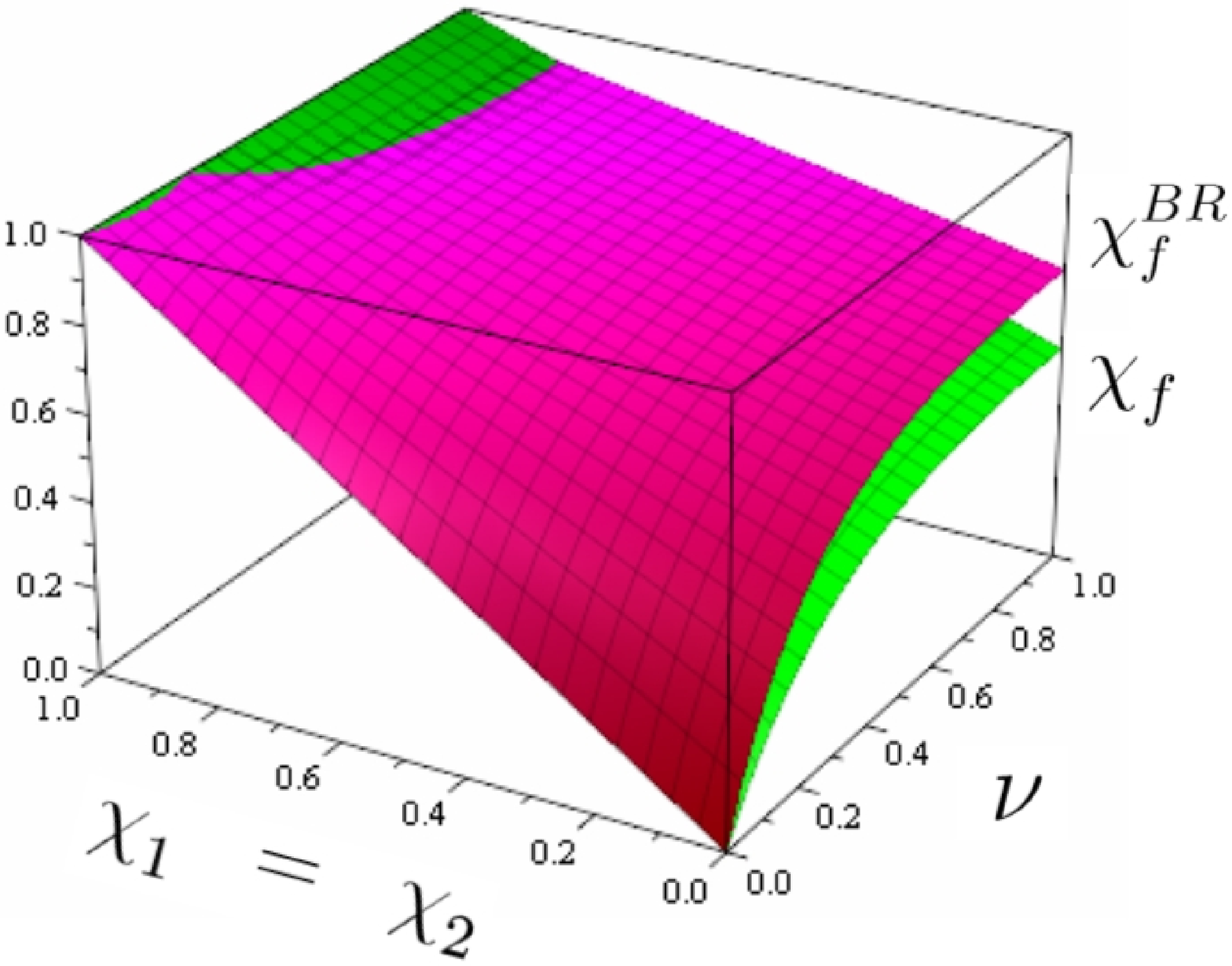}%
\includegraphics[width=7cm]{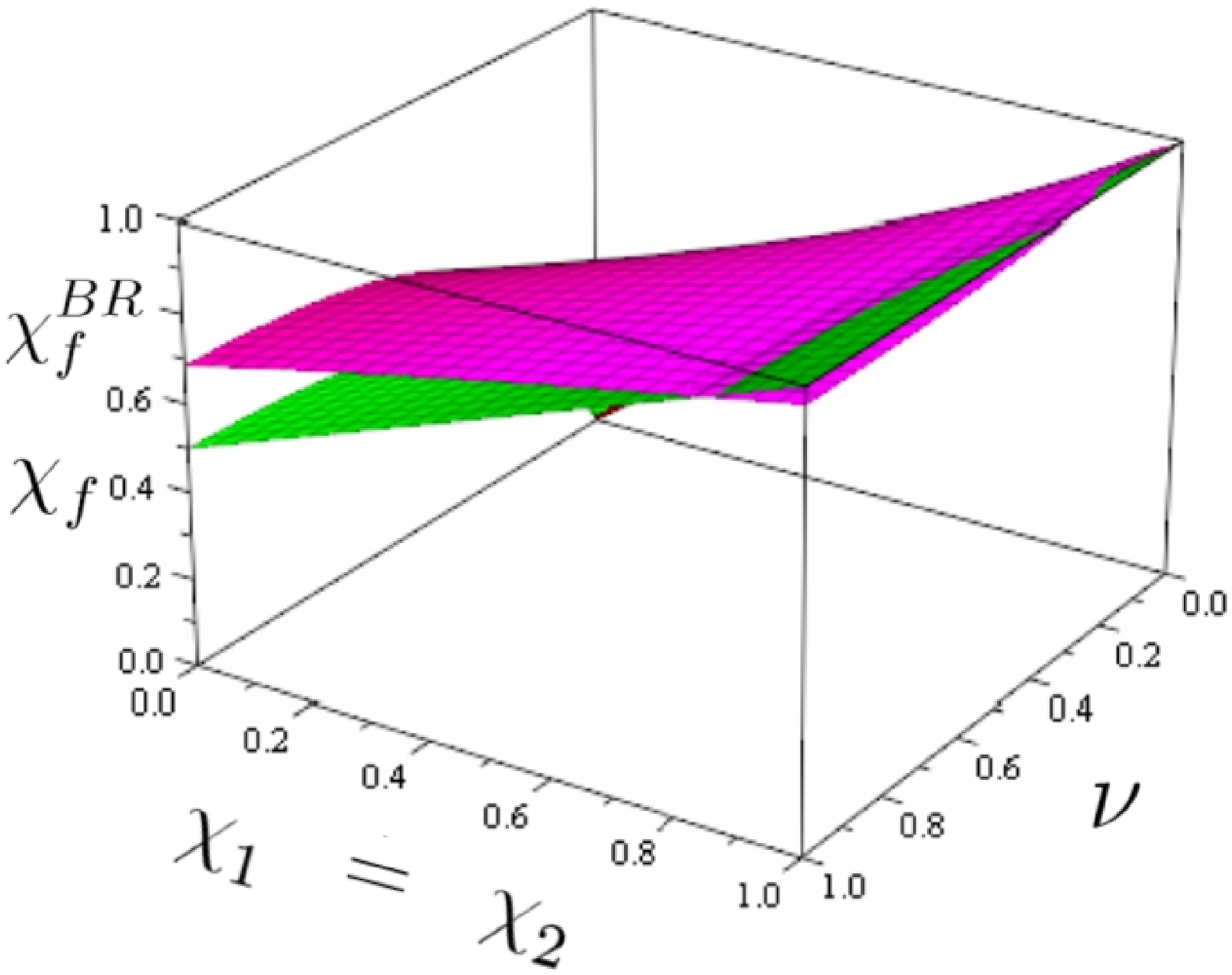} 
\includegraphics[width=7cm]{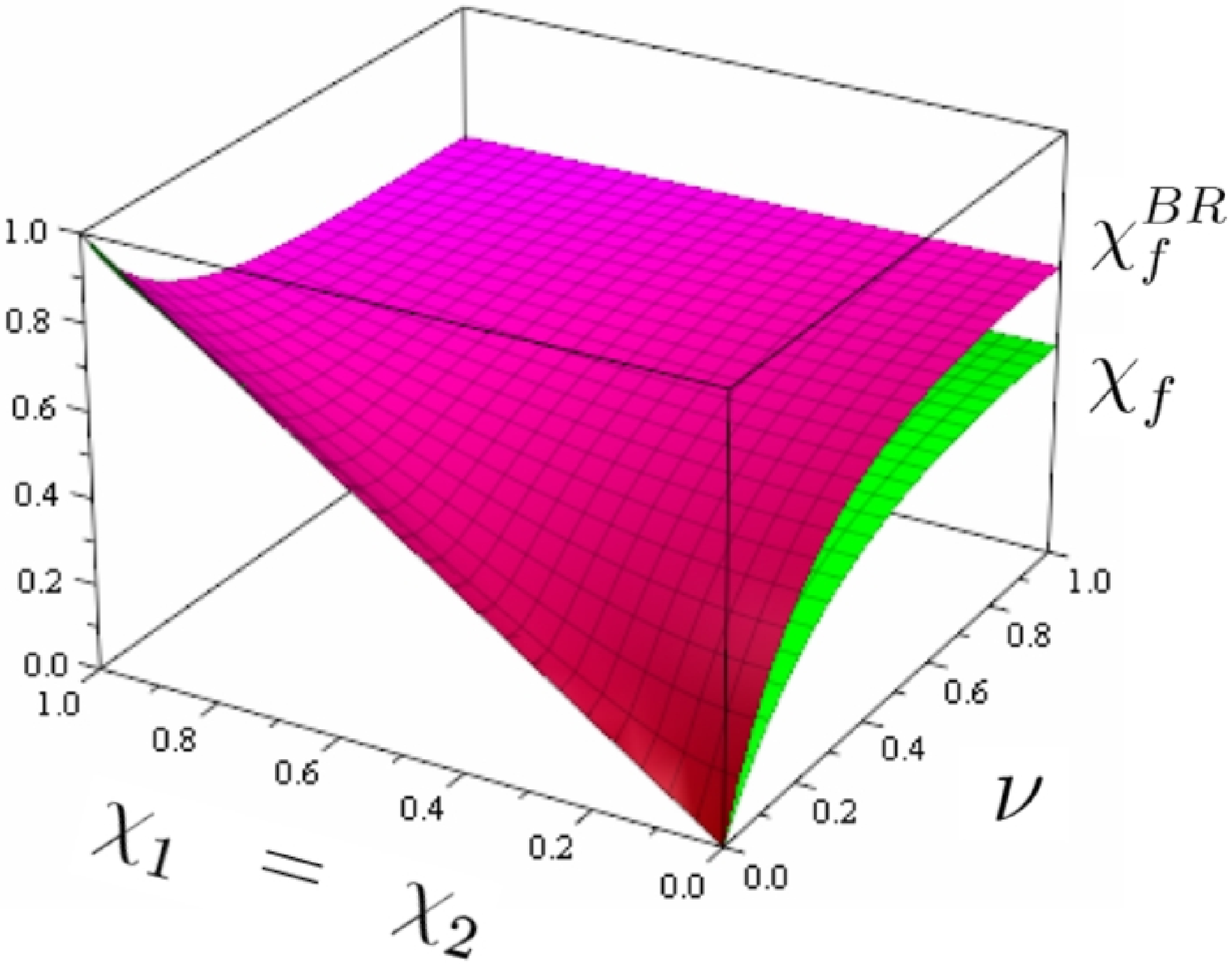}%
\includegraphics[width=7cm]{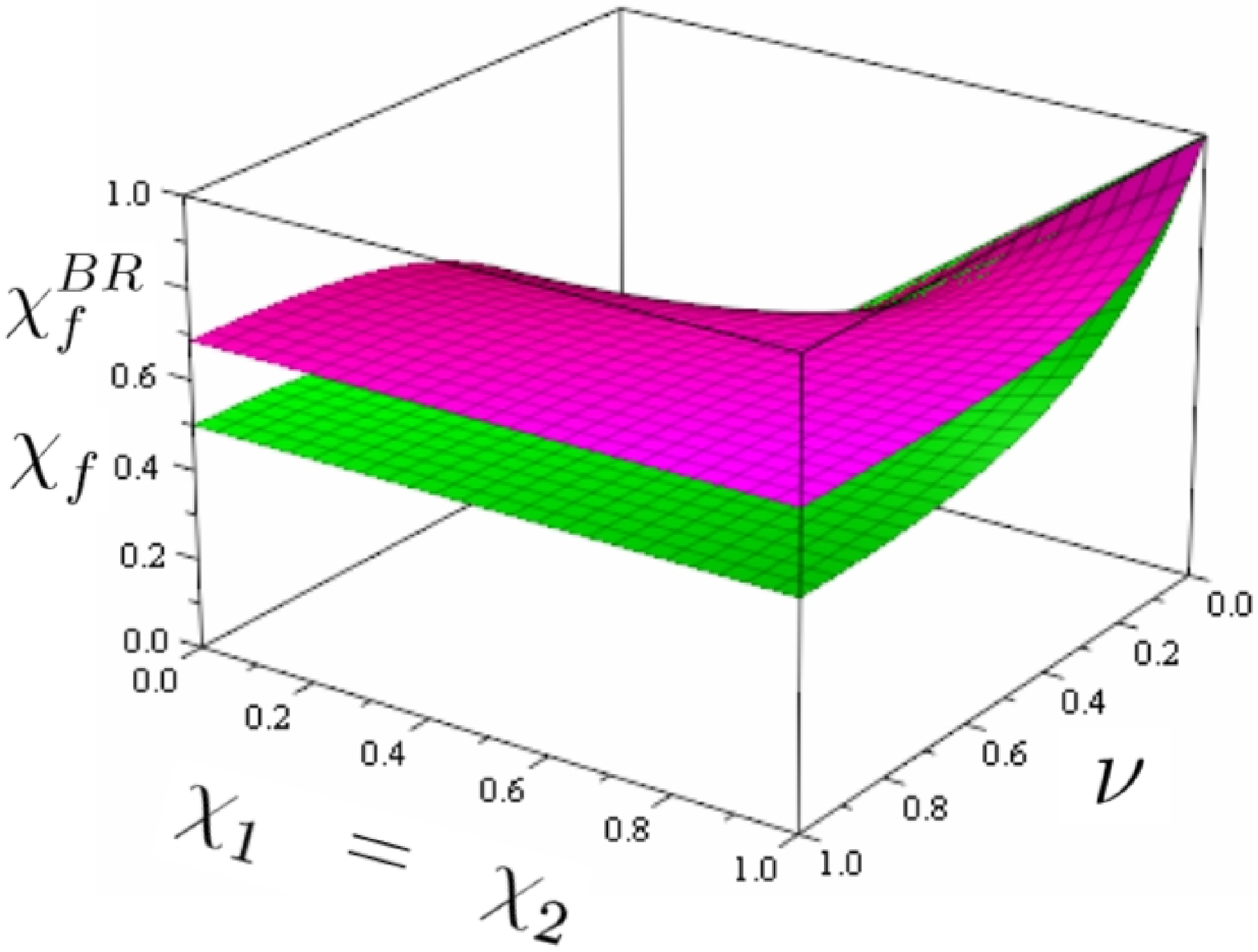}
\caption{(Color online)
The final spin estimates $\protect\chi _{f}$ (green surfaces) and 
$\protect\chi _{f}^{BR}$ (magenta) as function of $\protect\chi _{1}=\protect%
\chi _{2}$ and $\protect\nu $ for perfect alignment of the spins with the
orbital angular momentum (non-precessing case - upper row); and anti-aligned spins in the plane of
motion (severe precession - lower row). Except a narrow parameter range with high mass ratio and
high spin values in the upper row configuration, the estimated $\protect\chi %
_{f}$ is smaller than $\protect\chi _{f}^{BR}$. The agreement increases with
decreasing $\protect\nu $; for the aligned configuration (upper row) is
better in the high spin regime (visible on the right panel), then for low
spin (left).
}
\label{Fig1}
\end{figure}

\end{widetext}

\begin{widetext}

\begin{figure}[ht]
\includegraphics[width=7cm]{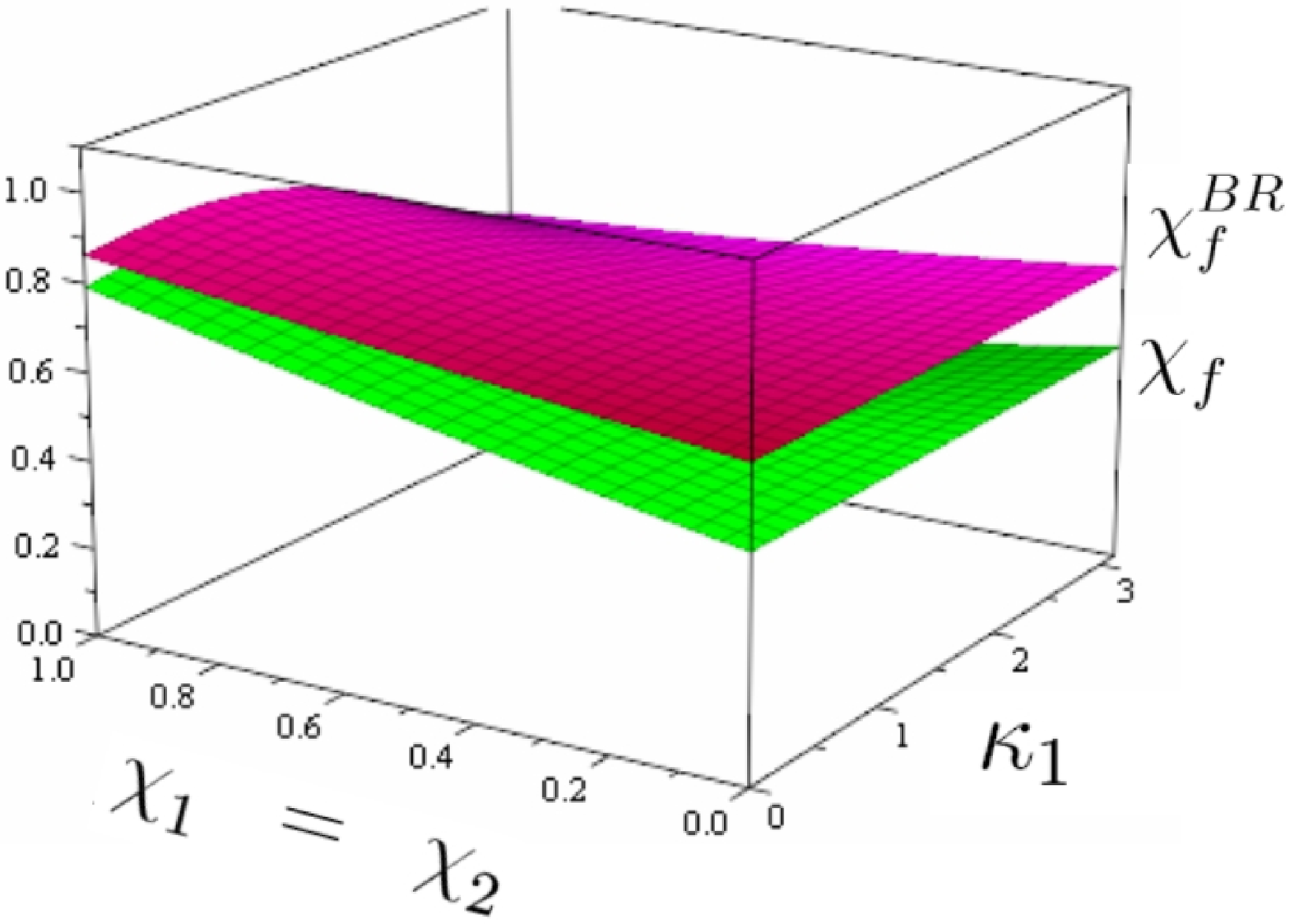}%
\includegraphics[width=7cm]{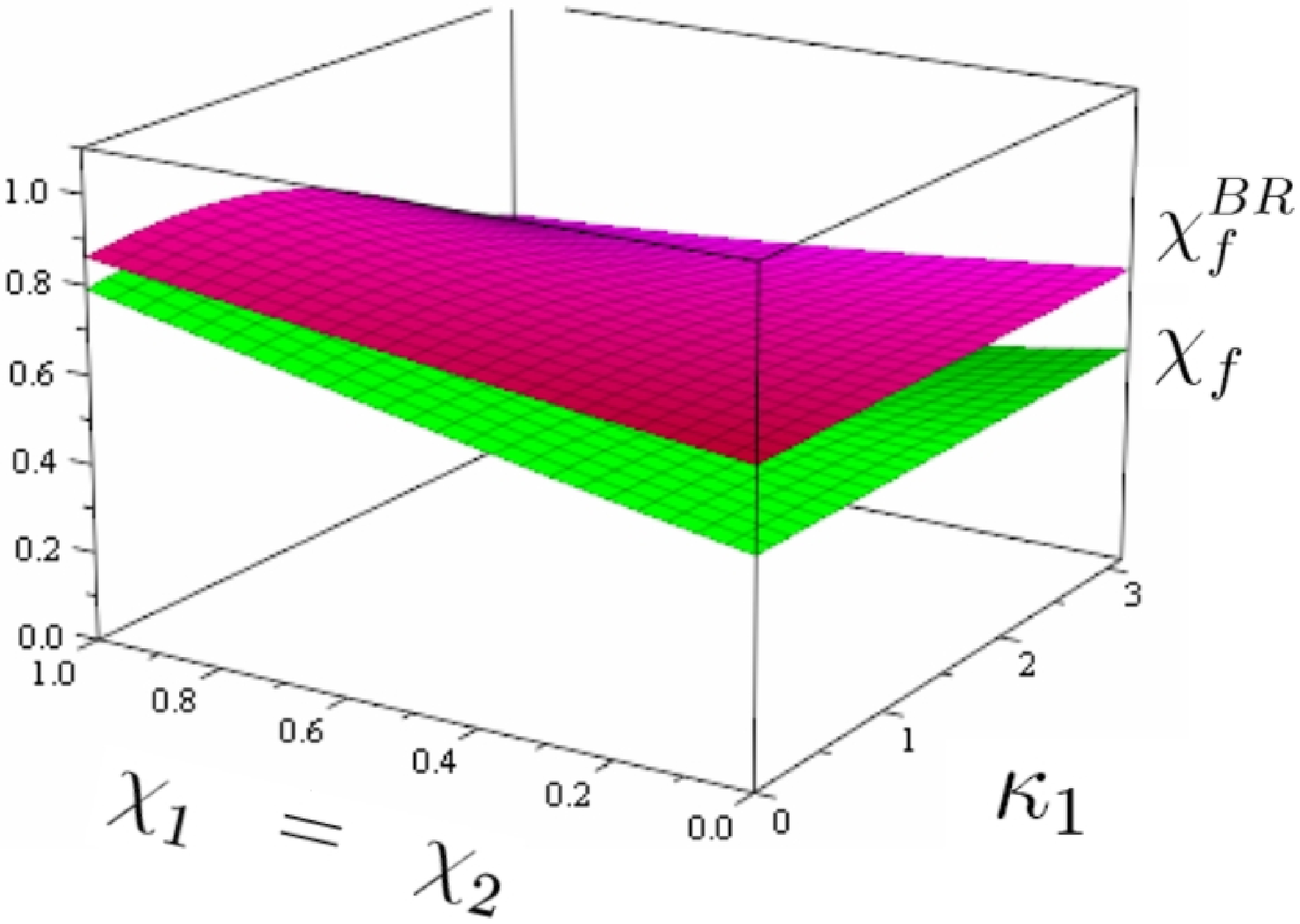} 
\includegraphics[width=7cm]{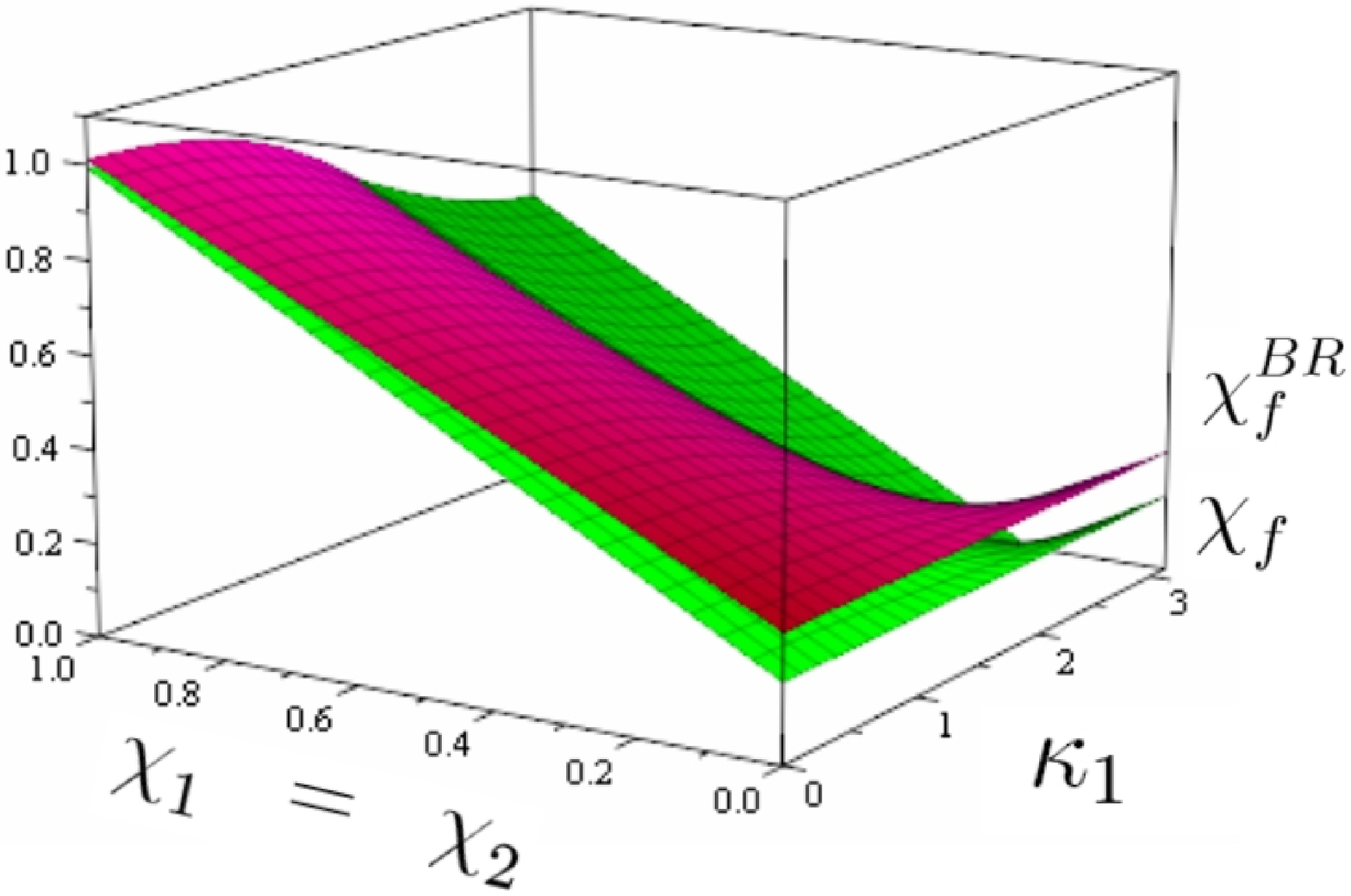}%
\includegraphics[width=7cm]{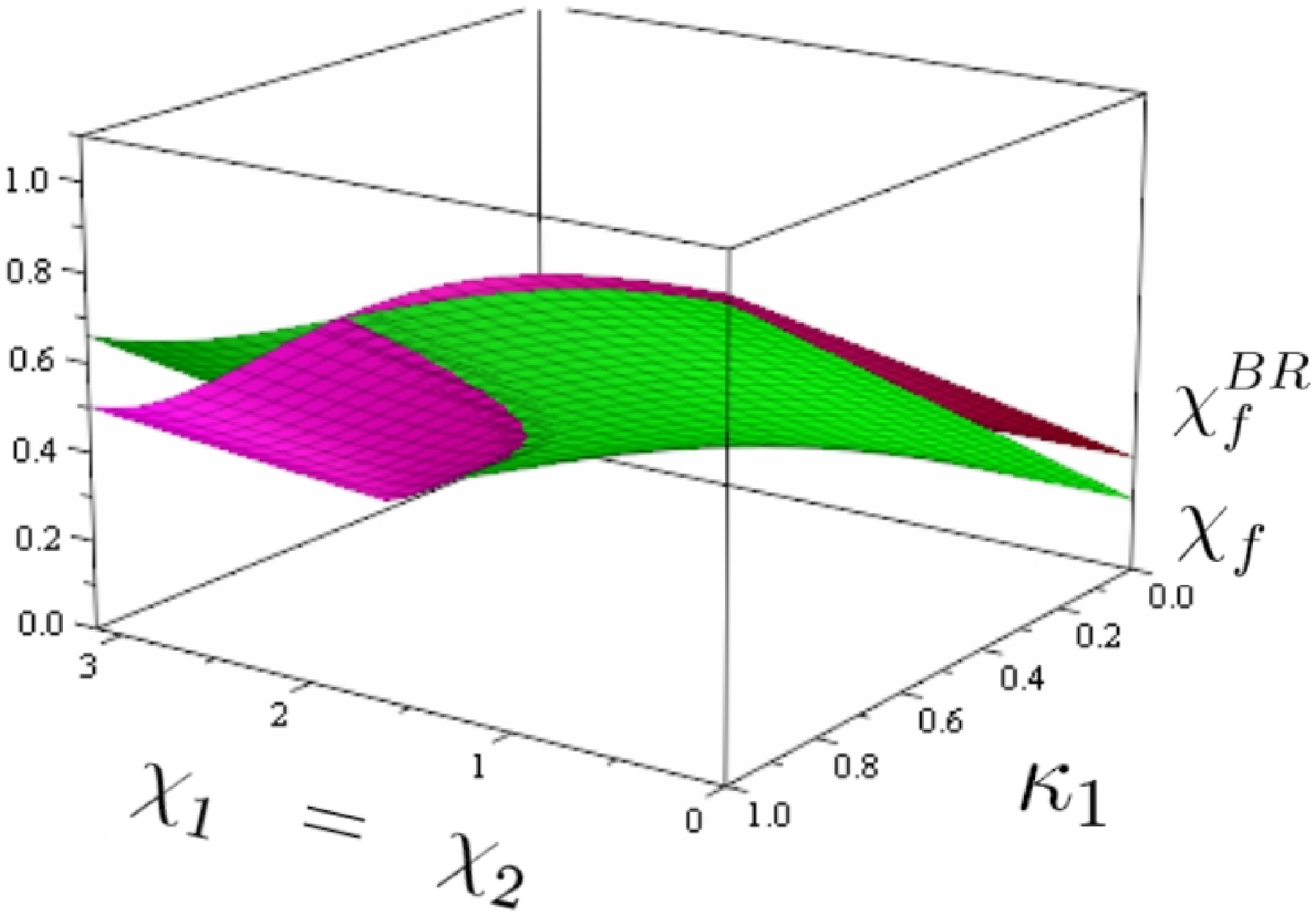} 
\includegraphics[width=7cm]{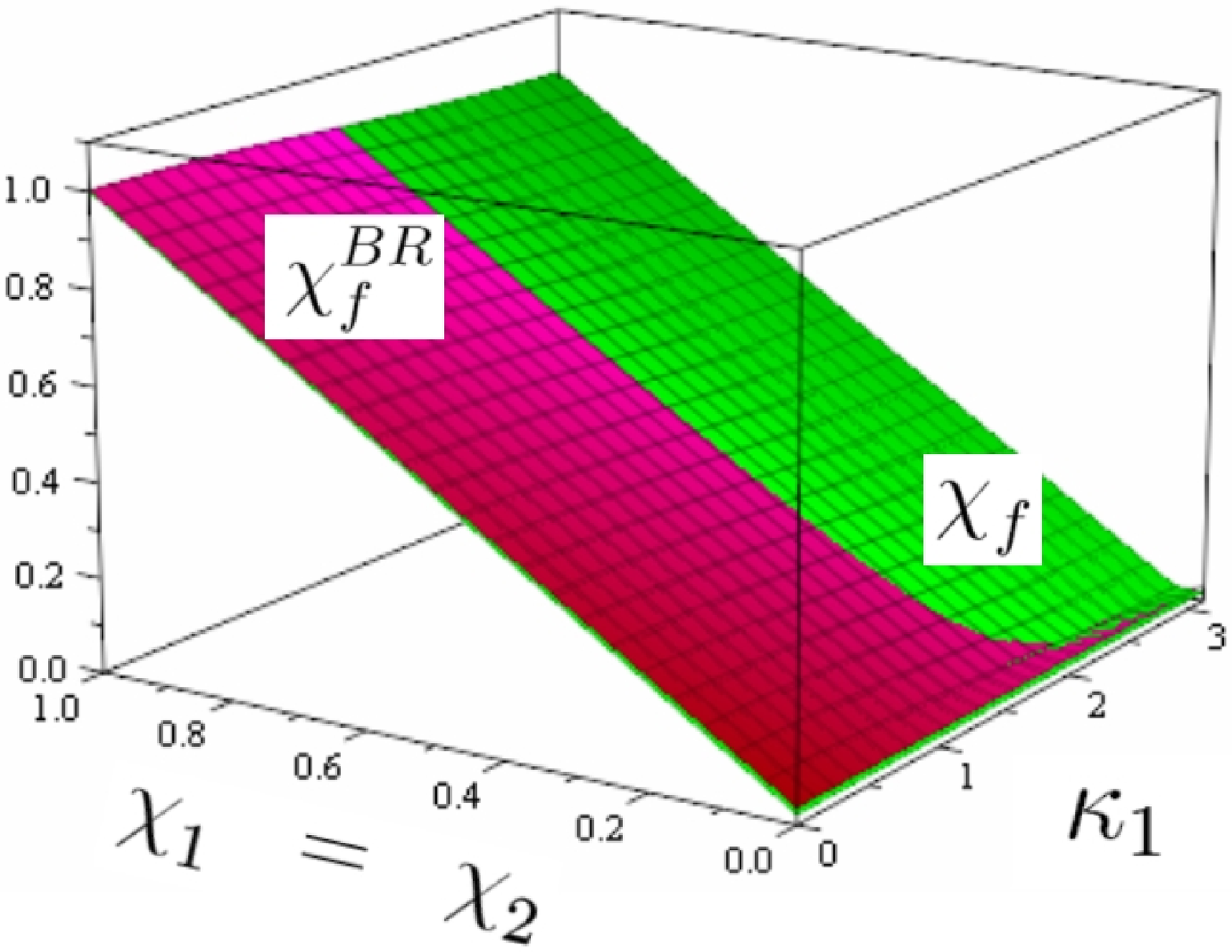}%
\includegraphics[width=7cm]{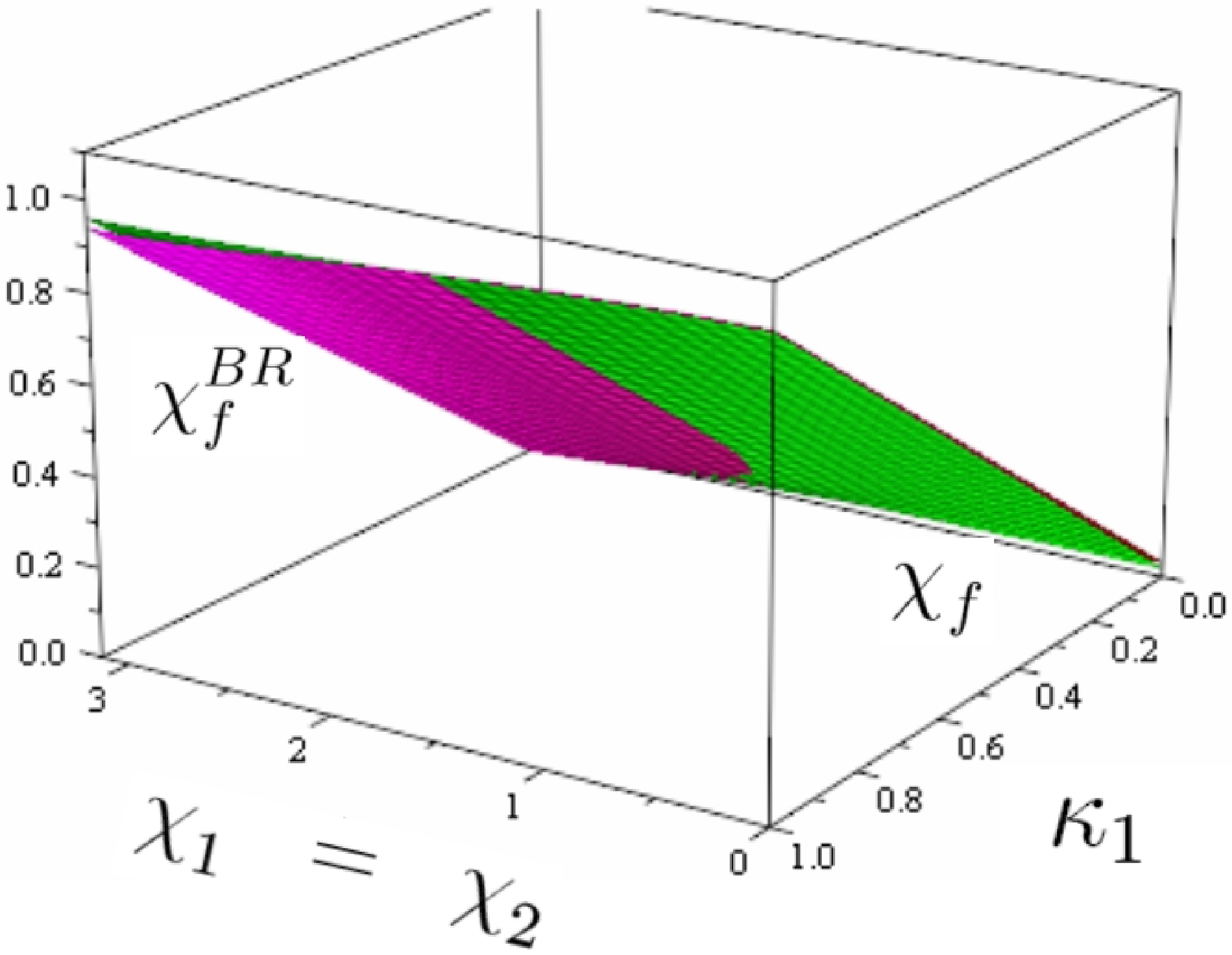}
\caption{(Color online) The final spin estimates $\protect\chi _{f}$ (green surfaces) and $%
\protect\chi _{f}^{BR}$ (magenta) as function of $\protect\chi _{1}=\protect%
\chi _{2}$ and $\protect\kappa _{1}$ for mass ratios $\protect\nu =1$ (upper
row), $\protect\nu =0.1$ (middle row) and $\protect\nu =0.01$ (lower row).
The represented configuration has the smaller spin confined to the plane
of motion and the larger spin lying in the plane span by the smaller spin
and orbital angular momentum. The agreement increases with decreasing $%
\protect\nu $ (with $\protect\chi _{f} <\protect\chi _{f} ^{BR}$ at large $\protect\nu 
$); is better in the high spin regime (visible on the right panel), then for
low spin (left); is also better for configurations with $\protect\kappa %
_{1}\in \left[ 0,\protect\pi /2\right] $ than for the severely misaligned
configurations.
}
\label{Fig2}
\end{figure}

\end{widetext}


\begin{thebibliography}{99}
\bibitem{LIGO} B. Abbott et al. (LIGO Scientific Collaboration), \textit{%
Rept. Prog. Phys.} \textbf{72}, 076901 (2009)
.

\bibitem{Virgo} F. Acernese et al., \textit{Class. Quantum Grav.} \textbf{25}%
, 184001 (2008).

\bibitem{GEO} H. Grote and the LIGO Scientific Collaboration, \textit{Class.
Quantum Grav.} \textbf{25}, 114043 (2008).

\bibitem{ET} J. R. Gair, I. Mandel, M. C. Miller, M. Volonteri,\ \textit{%
Gen. Relativ. Gravit.} \textbf{43} 485 (2011).

\bibitem{IMBH} S. A. Farrell, N. A. Webb, D. Barret, O. Godet, J. M.
Rodrigues, \textit{Nature} \textbf{460}, 73 (2009); M. Mezcua, A. P.
Lobanov, \textit{Compact radio emission in Ultra Luminous X-ray sources},
E-print: arXiv:1011.0946, Proceedings of the conference "Ultra-Luminous
X-ray sources and Middle Weight Black Holes", to appear in \textit{%
Astronomische Nachrichten}.

\bibitem{Kormendy} J. Kormendy, N. Drory, R. Bender, M. E. Cornell, \textit{%
Astrophys. J.} \textbf{723}, 54 (2010)%
.

\bibitem{Bardeen} J. M. Bardeen, \textit{Nature} \textbf{226}, 64 (1970).

\bibitem{PageThorne} D. N. Page, K. S. Thorne, \textit{Astrophys. J.} 
\textbf{191}, 499 (1974)

\bibitem{accretion} M. Abramowicz, M. Jaroszynski, M. Sikora, \textit{%
Astron. Astrophys.}, \textbf{63}, 221 (1978); R. D. Blandford, R. L. Znajek, 
\textit{Month. Not. Roy. Astr. Soc. }\textbf{179}, 433 (1977); M. Camenzind, 
\textit{Astron. \& Astroph.} \textbf{156}, 137 (1986); M. Camenzind, \textit{%
Astron. \& Astroph.} \textbf{162}, 32 (1986); M. Camenzind, \textit{Astron.
\& Astroph.} \textbf{184}, 341 (1987); M. Takahashi, S. Nitta, Y. Tamematsu,
A. Tomimatsu, \textit{Astrophys. J.} \textbf{363}, 206 (1990); S. Y. Nitta,
M. Takahashi, A. Tomimatsu, \textit{Phys. Rev. D} \textbf{44}, 2295 (1991);
K. Hirotani, M. Takahashi, S. Nitta, A. Tomimatsu, \textit{Astrophys. J.} 
\textbf{386}, 455 (1992); H. Falcke, P. L. Biermann, \textit{Astron. \&
Astroph.} \textbf{293}, 665 (1995); L. X. Li, \textit{Astrophys. J.} \textbf{%
533}, L115 (2000); D. X. Wang, K. Xiao, W. H. Lei, \textit{Month. Not. Roy.
Astr. Soc.} \textbf{335}, 655 (2002); D. A. Uzdensky, \textit{Astrophys. J.} 
\textbf{620}, 889 (2005); Z. Kov\'{a}cs, P. L. Biermann, L. \'{A}. Gergely, 
\textit{Mon. Not. Royal Astron. Soc.} \textbf{416}, 991 (2011); Z. Kov\'{a}%
cs, L. \'{A}. Gergely, M Vas\'{u}th, \textit{Phys Rev. }D \textbf{84},
024018 (2011).

\bibitem{cutoff} J. P. Leahy, T. W. B. Muxlow, P. W. Stephens, \textit{%
Monthly Not. Royal Astron. Soc.} \textbf{239}, 401 (1989)%
; C. L. Carilli, R. A. Perley, J. W. Dreher, J. P. Leahy, \textit{Astrophys.
J.} \textbf{383}, 554 (1991)%
; A. Celotti, A. C. Fabian, \textit{Monthly Not. Royal Astron. Soc.} \textbf{%
264}, 228 (1993)%
; W. J. Duschl, H. Lesch, \textit{Astron. Astrophys.} \textbf{286}, 431
(1994)
.

\bibitem{UHECR} P. L. Biermann, J. K. Becker, L. Caramete, L. \'{A}.
Gergely, I. C. Mari\c{s}, A. Meli, V. de Souza, T. Stanev, \textit{Int. J.
Mod. Phys.} D \textbf{18} 1577 (2009); P. L. Biermann, J. K. Becker, L.
Caramete, A. Curutiu, R. Engel, H. Falcke, L. \'{A}. Gergely, P. G. Isar, I.
C. Mari\c{s}, A. Meli, K.-H. Kampert, T. Stanev, O. Ta\c{s}c\u{a}u, C. Zier, 
\textit{Nucl. Phys. B, Proc. Suppl.} \textbf{190}, 61 (2009); Gopal-Krishna,
P. L. Biermann, V. de Souza, P. J. Wiita, \textit{Astrophys. J. Lett.} 
\textbf{720}, L155 (2010); P. L. Biermann, V. de Souza, \textit{Astrophys. J.%
} \textbf{746}, 72 (2012).

\bibitem{Zier} Ch. Zier, \textit{Monthly Not. Royal Astron. Soc. Lett.} 
\textbf{371}, L36 (2006).

\bibitem{Alexander} T. Alexander, in \textit{2007 STScI Spring Symp.: Black
Holes}, ed. M. Livio \& A. M. Koekemoer (Cambridge: Cambridge Univ. Press),
E-print: arXiv:0708.0688.

\bibitem{Sesana} A. Sesana, F. Haardt, P. Madau, \textit{Astrophys. J.} 
\textbf{651}, 392 (2006); A. Sesana, F. Haardt, P. Madau,, \textit{%
Astrophys. J.} \textbf{660}, 546 (2007); A. Sesana, F. Haardt, P. Madau,, 
\textit{Astrophys. J.} in press (2007), arXiv:0710.4301.

\bibitem{Hayasaki} K. Hayasaki, \textit{PASJ} (2008), E-print:
arXiv:0805.3408.

\bibitem{SpinFlip1} L. \'{A}. Gergely, P. L. Biermann, \textit{Astrophys. J.}
\textbf{697}, 1621 (2009).

\bibitem{LISA} K. G. Arun, S. Babak, E. Berti, N. Cornish, C. Cutler, J.
Gair, S. A. Hughes, B. R. Iyer, R. N. Lang, I. Mandel, E. K. Porter, B. S.
Sathyaprakash, S. Sinha, A. M. Sintes, M. Trias, C. Van Den Broeck, M.
Volonteri, \textit{Class. Quantum Grav.} \textbf{26}, 094027 (2009); R. N.
Lang, S. A. Hughes, \textit{Class. Quantum Grav.} \textbf{26}, 094035 (2009).

\bibitem{BaPe} J. M. Bardeen, J. A. Petterson, \textit{Astrophys. J.} 
\textbf{195}, L65 (1975)%
.

\bibitem{HuBl} S. A. Hughes, R. D. Blandford, \textit{Astrophys. J.} \textbf{%
585}, L101 (2003).

\bibitem{BertiVolonteri} E. Berti, M. Volonteri, \textit{Astrophys. J.} 
\textbf{684}, 822 (2008).

\bibitem{SpinFlip2} L. \'{A}. Gergely, P. L. Biermann, L. I. Caramete, 
\textit{Class. Quantum Grav.} \textbf{27}, 194009 (2010).

\bibitem{CarameteBiermann} L. I. Caramete, P. L. Biermann, \textit{Astron.
Astroph.} \textbf{521}, A55 (2010).

\bibitem{PressSchechter} W. H. Press, P. Schechter, \textit{Astrophys. J.} 
\textbf{187}, 425 (1974).

\bibitem{WilsonColbert} A. S. Wilson, E. J. M. Colbert, \textit{Astrophys. J.%
} \textbf{438}, 62 (1995).

\bibitem{Lauer} T. R. Lauer et al., \textit{Astrophys. J.} \textbf{662},
808L (2007).

\bibitem{Ferrarese2} L. Ferrarese et al., \textit{Astrophys. J. Suppl. Ser.} 
\textbf{164}, 334 (2006).

\bibitem{Cote} P. C\^{o}t\'{e}, S. Piatek, L. Ferrarese, et al., \textit{%
Astrophys. J. Suppl. Ser.} \textbf{165}, 57 (2006).

\bibitem{Magorrian} J. Magorrian et al., \textit{Astronomical J.} \textbf{115%
}, 2285 (1998).

\bibitem{Benson} A. J. Benson, D. D\v{z}anovi\'{c}, C. S. Frenk, R.
Sharples, \textit{Mon. Not. Roy. Astron. Soc.} \textbf{379}, 841 (2007).

\bibitem{Gilmore} G. Gilmore et al., \textit{Nucl. Phys. }B \textbf{173}, 15
(2007).

\bibitem{Klypin} A. Klypin, H.-S. Zhao, R. S. Somerville, \textit{Astrophys.
J.} \textbf{573}, 597 (2002).

\bibitem{BOC} B. M. Barker, R. F. O'Connell, \textit{Phys. Rev. D} \textbf{12%
}, 329 (1975); B. M. Barker, R. F. O'Connell, \textit{Gen. Relativ. Gravit.} 
\textbf{11}, 149 (1979).

\bibitem{ACST} A. Apostolatos, C. Cutler, G. J. Sussman, K. S. Thorne, 
\textit{Phys. Rev. D }\textbf{49}, 6274 (1994).

\bibitem{GPV3} L. \'{A}. Gergely, Z. I. Perj\'{e}s, M. Vas\'{u}th, \textit{%
Phys. Rev.} D \textbf{58}, 124001 (1998).

\bibitem{spinspin1} L. \'{A}. Gergely, \textit{Phys. Rev.} D \textbf{61},
024035 (1999).

\bibitem{quadrup} L. \'{A}. Gergely, Z. Keresztes, \textit{Phys. Rev.} D 
\textbf{67}, 024020 (2003).

\bibitem{spinspin2} L. \'{A}. Gergely, \textit{Phys. Rev.} D \textbf{62},
024007 (2000).

\bibitem{finalspin} L. Rezzolla, E. Barausse, E. Nils Dorband, D. Pollney,
C. Reisswig, J. Seiler, S. Husa, \textit{Phys. Rev.} D \textbf{78} 044002
(2008); M. C. Washik , J. Healy, F. Herrmann, I. Hinder, D. M. Shoemaker, P.
Laguna, R. A. Matzner, \textit{Phys. Rev. Lett.} \textbf{101} 061102 (2008);
A. Buonanno, L. E. Kidder, L. Lehner, \textit{Phys. Rev.} D \textbf{77}
026004 (2008); W. Tichy, P. Marronetti, \textit{Phys. Rev.} D \textbf{78},
081501(R) (2008); U. Sperhake, V. Cardoso, F. Pretorius, E. Berti, T.
Hinderer, N. Yunes, \textit{Phys. Rev. Lett.} \textbf{103}, 131102 (2009);
J. Healy, P. Laguna, R. A. Matzner, D. M. Shoemaker, \textit{Phys. Rev.} D 
\textbf{81}, 081501 (2010)%
; E. Barausse, \textit{The importance of precession in modelling the
direction of the final spin from a black-hole merger,} E-print:
arXiv:0911.1274 (2009); M. Kesden, U. Sperhake, E. Berti, \textit{Phys. Rev. 
}D \textbf{81}, 084054 (2010)
; C. O. Lousto, M. Campanelli, Y. Zlochower, H. Nakano, \textit{Class.
Quantum Grav.} \textbf{27}, 114006 (2010)%
.

\bibitem{BR} E. Barausse, L. Rezzolla, \textit{Astrophys. J. Lett.} \textbf{%
704} L40-L44 (2009).

\bibitem{Greene} J. E. Greene, A. J. Barth, L. C. Ho, \textit{New Astron.
Rev.} \textbf{50}, 739 (2006); J. E. Greene, L. C. Ho, \textit{Astrophys. J.}%
.\textbf{641}, L21 (2006).

\bibitem{Inspiral} L. \'{A}. Gergely, \textit{Phys. Rev.} D \textbf{81},
084025 (2010); L. \'{A}. Gergely, \textit{Phys. Rev. }D \textbf{82,} 104031
(2010).
\end{thebibliography}
\end{document}